%% file: main.tex
\documentclass[lettersize,journal]{IEEEtran}
\usepackage{amsmath,amsthm,amssymb,amsfonts}
\usepackage{algorithmic}
\usepackage{algorithm}
\usepackage{array}
\usepackage{textcomp}
\usepackage{stfloats}
\usepackage{url}
\usepackage{verbatim}
\usepackage{graphicx}
\usepackage{cite}
\usepackage{booktabs}
\usepackage{multirow} 
\usepackage{xcolor}
\usepackage{diagbox}
\usepackage{tabularray}
\usepackage{caption}
\usepackage{subcaption}
\usepackage{float} 
\usepackage{graphicx}
\usepackage{hyperref}
\hypersetup{breaklinks=true}
\usepackage{authblk}
\usepackage[normalem]{ulem}
\usepackage{tikz}
\usepackage{tikz-cd}
\usetikzlibrary{arrows.meta}
\usepackage{color}
\usepackage{marvosym} 
\newcommand{\corremail}{\href{mailto:xinran.zheng.23@ucl.ac.uk}{\textsuperscript{\normalfont\Letter}}}

\begin{document}
\newcommand{\model}{{\textsf{TIF}}}
\newtheorem{myDef}{Definition}
\raggedbottom
\newcommand{\diff}[1]{\textcolor{black}{#1}}


\title{\model: Learning Temporal Invariance in Android Malware Detectors}



\author{Xinran Zheng\IEEEauthorrefmark{1}\corremail, 
Shuo Yang\IEEEauthorrefmark{1}, Edith C.-H. Ngai, Suman Jana and Lorenzo Cavallaro%
\thanks{Xinran Zheng and Lorenzo Cavallaro are with University College London, London, UK. Shuo Yang and Edith C.-H. Ngai are with The University of Hong Kong, Hong Kong SAR, China. Suman Jana is with Columbia University, New York, USA.}%
\thanks{\IEEEauthorrefmark{1}Equal Contribution}%
}



\markboth{IEEE Transactions on Software Engineering}%
{Shell \MakeLowercase{\textit{et al.}}: A Sample Article Using IEEEtran.cls for IEEE Journals}


\maketitle

\begin{abstract}
Learning-based Android malware detectors degrade over time due to natural distribution drift caused by malware variants and new families. This paper systematically investigates the challenges classifiers trained with empirical risk minimization (ERM) face against such distribution shifts and attributes their shortcomings to their inability to learn \emph{stable} discriminative features. Invariant learning theory offers a promising solution by encouraging models to generate stable representations across environments that expose the instability of the training set. However, the lack of prior environment labels, the diversity of drift factors, and low-quality representations caused by diverse families make this task challenging. To address these issues, we propose \model, the first temporal invariant training framework for malware detection, which aims to enhance the ability of detectors to learn stable representations across time. \model~organizes environments based on application observation dates to reveal temporal drift, integrating specialized multi-proxy contrastive learning and invariant gradient alignment to generate and align environments with high-quality, stable representations. \model~can be seamlessly integrated into any learning-based detector. Experiments on a decade-long dataset show that \model~excels, particularly in early deployment stages, addressing real-world needs and outperforming state-of-the-art methods.
\end{abstract}


\input{Tex/Introduction}
\input{Tex/Review}
\input{Tex/Motivation}
\input{Tex/Methodology}
\input{Tex/Experiment}

\input{Tex/Discussion}
\input{Tex/Conclusion}
\bibliographystyle{IEEEtran}
\bibliography{reference}
\appendix
\input{Tex/Appendix}

\end{document}

%% file: Tex/Introduction.tex
\section{Introduction}
\label{sec:introduction}
In open and dynamic environments, even the most effective malware detectors encounter significant challenges due to natural distribution drift, leading to performance degradation~\cite{transcending, cade}. This degradation arises from the continuous evolution of malware behavior and the emergence of new families that detectors have not previously encountered~\cite{pei2024exploiting}. These new variants alter the underlying statistical properties of test samples~\cite{tesseract, overkill, transcending, Drift_forensice}, thereby weakening detectors that rely on features derived from past training data.


To understand the rapid degradation of detectors under natural distribution drift, we revisit the standard training paradigm and attribute this vulnerability primarily to the failure to capture \textit{temporal invariant semantics}. We observe that semantic invariance exists both within and across malware families; while implementation details (e.g., APIs or control flows) may change, the underlying malicious intent remains relatively stable. The failure to capture such semantics stems from the inherent limitation of Empirical Risk Minimization (ERM), the dominant machine learning training paradigm. ERM implicitly assumes all training samples are drawn from a single stationary distribution. By flattening malware instances collected at different times into a unified optimization objective, ERM treats all correlations homogeneously. This encourages the model to indiscriminately latch onto statistical shortcuts that are predictive on average but unstable over time. Consequently, the learned detector overfits to transient patterns, disregarding the invariant semantics implicitly present in historical data.

Recent research has focused on diagnosing the root causes of drift, yet these approaches are inherently limited. Some methods explicitly attribute the dynamic nature of malware to specific causes, such as unseen families~\cite{cade, DOMR} or API renaming~\cite{apigraph}. Others leverage observational heuristics of data dynamics, such as the distinct learning pacing of robust versus unstable correlations~\cite{scrr} or the monotonic evolution of feature importance over time~\cite{svm_ce}. However, natural drift is driven by multifaceted factors, where features often exhibit complex, non-linear dependencies. Simply targeting isolated drift sources yields only partial alleviation and often fails to generalize beyond specific target settings. Alternatively, Continual Learning (CL) addresses drift by continuously updating models with new data, without explicitly reasoning about the underlying causes~\cite{continuous, Temporal-incremental}. Yet, maintaining such systems is costly due to the constant need for expert annotation or the risk of noise from pseudo-labeling~\cite{labelless}. Moreover, performance degradation between updates undermines model robustness, compromising the reliability of continuous deployment. Taken together, these limitations underscore a central research question: whether invariant malicious semantics already embedded in historical training data can be explicitly exploited to address the limitations of ERM under natural distribution drift, without relying on drift factor modeling or frequent model updates.

Invariant learning theory~\cite{IR_intro} aims to address the shortcomings of ERM by learning invariant features/representations shared across different distributions, which aligns with our objective. It promotes the discovery of stable representations by dividing training data into distinct subsets or ``environments'' and encourages the model to minimize differences between them to capture stable elements across environments. This is based on two premises: a priori environment labels that reveal instability~\cite{environment_label, env_label} and high-quality representations that adequately encode feature information~\cite{yang2024invariant}. These assumptions are not trivial for malware detection, as malware evolution is attributed to a variety of non-obvious factors, and the extremely unbalanced sample distributions and complex feature spaces due to multiple malicious families increase the complexity of learning high-quality representations, making invariance across environments difficult to explore.

In this paper, we present a \textbf{T}emporal \textbf{I}nvariant Training \textbf{F}ramework (\model) to promote invariant representation learning under natural distribution drift. 
\model~partitions training data temporally based on application observation dates, treating different time periods as distinct environments without relying on predefined environment labels. 
To support invariant learning in binary malware detection with heterogeneous multi-family malware samples, we introduce a multi-proxy contrastive learning module that facilitates modular and expressive representation learning within each class. 
Moreover, \model~adopts a tailored invariant optimization strategy that encourages consistent optimization behavior for samples of the same class across temporal environments, thereby enhancing temporally invariant representations. 
Our framework is orthogonal to existing robust malware detectors: it requires no changes to feature spaces or model architectures and can be integrated as an enhanced training paradigm for learning-based detectors. 
The main contributions of this paper are as follows:

\begin{itemize}
    \item We formalize malware evolution under natural distribution drift from an invariant learning perspective, identifying discriminative and stable invariant representations as key to robust malware detection (Section~\ref{motivation}).
    \item We design a multi-proxy contrastive learning module (Section~\ref{multi-proxy contrastive learning}) to model heterogeneous multi-family malware distributions and an invariant gradient alignment module (Section~\ref{invariant_alignment}) to enforce consistent optimization across temporal environments, promoting high-quality stable representations.
    \item We present \model, a temporal invariant training framework that integrates with arbitrary learning-based detectors and feature spaces, enabling robust representation learning by exposing and suppressing drift-induced unstable information (Section~\ref{invariant training}).
    \item We construct a 10-year dataset and experiments to evaluate \model's robustness across drift scenarios and feature spaces. Results show that \model~slows degradation and outperforms state-of-the-art methods (Section~\ref{evaluation}).
\end{itemize}

%% file: Tex/Review.tex
\section{Background}
In this section, we review drift-robust malware detectors and the key components of invariant learning, motivating a temporal invariant representation learning solution for Android malware detection.

\subsection{Natural Drift of Malware}
\diff{Natural drift refers to temporal changes in data distributions caused by malware evolution or emerging trends. 
Such drift violates the i.i.d. assumption and degrades malware detectors by shifting decision boundaries over time. 
In Android malware, two primary sources are commonly identified. 
First, malware may evolve within the same family due to system updates, framework changes, or advanced obfuscation techniques~\cite{malware_evolution_obf,xue2021happer,aghakhani2020malware}, altering feature distributions and making models trained on earlier data outdated~\cite{malware_evolution_update}. 
Second, new malware families may emerge during deployment, exhibiting patterns unseen during training~\cite{new_family} and leading to unreliable predictions.}

\subsection{Drift-robust Malware Detectors}
\label{Drift-robust Malware Detectors}

\diff{
Existing drift-robust malware detectors can be broadly grouped into two categories. The first line of work improves robustness by explicitly modeling or attributing drift to specific causes. 
Representative methods assume that drift mainly arises from API renaming or updating~\cite{apigraph}, the emergence of unseen malware families~\cite{DOMR}, or observable training dynamics~\cite{scrr, svm_ce}. 
For example, Yang \textit{et al.}~\cite{scrr} exploit the observation that spurious correlations are often learned faster than robust ones, and identify unstable features through learning-speed disparities. 
Angioni \textit{et al.}~\cite{svm_ce} assume that feature importance changes monotonically over time and suppress features whose contribution varies significantly across temporal intervals, which is mainly applicable to linear classifiers. 
By targeting a specific drift factor or proxy, these methods identify and filter unstable features accordingly. 
However, real-world malware evolution is driven by complex, multi-factor interactions that are difficult to isolate~\cite{transcend}. 
Heuristics based on a single assumed cause or manifestation may fail to capture the full spectrum of instability when multiple drift sources coexist.}

\diff{
The other category addresses drift through incremental or continual learning~\cite{continuous, Temporal-incremental}, where detectors are periodically updated with newly labeled drifted samples from deployment. 
Although extending model lifetime, this paradigm introduces a performance-cost trade-off, requiring sustained annotation and computation. 
Moreover, it adapts models to evolving distributions without explaining why learned representations fail to remain stable, so performance may still degrade between updates in long-term deployment.}

In summary, prior drift-robust malware detectors either rely on rigid assumptions about the causes of drift, or compensate for drift by repeatedly updating models with new data.
Given the inevitability and complexity of natural malware evolution, predefining what should remain stable is inherently challenging.
This motivates a different perspective that designs learning objectives that explicitly encourage the extraction of drift-robust representations without assumptions about temporal drift causes and without relying on frequent retraining.

\subsection{Invariant Learning}
\label{invariant_learning}
Assume that the training data $\mathcal{D}_{tr}$ are collected from multiple environments $e \in \mathcal{E}$, i.e., $\mathcal{D}_{tr}=\{D^e_{tr}\}_{e \in \mathcal{E}}$. 
Let the input and target spaces be $x \in \mathcal{X}$ and $y \in \mathcal{Y}$, respectively. 
For observations $x,y \sim p(x,y|e)$ from each environment, samples within an environment are independently and identically distributed. 
Consider a classification model $f=c \circ \phi$, where $\phi:\mathcal{X} \rightarrow \mathcal{H}$ maps inputs into the representation space $\mathcal{H}$, and $c:\mathcal{H} \rightarrow \mathcal{Y}$ maps representations to logits over $\mathcal{Y}$.

\subsubsection{Learning Invariant Representation}
\label{Learning Invariant Representation}
\diff{Under distribution drift, test data $\mathcal{D}_{te}$ may be drawn from an unseen environment $e_{te} \notin \mathcal{E}$, requiring models to generalize beyond training distributions. 
Invariant learning addresses this challenge by learning representations whose predictive relationship with the label remains stable across environments. 
Formally, it seeks a classifier $c(\cdot)$ that satisfies the environmental invariance constraint (EIC)~\cite{EIC}:
\begin{equation}
\mathbb{E}\left[y \mid \phi(x), e\right] = \mathbb{E}\left[y \mid \phi(x), e^{\prime}\right], \quad \forall e, e^{\prime} \in \mathcal{E}.
\end{equation}}
\diff{This constraint is typically imposed through a penalty term:
\begin{equation}
\min _f\sum_{e \in \mathcal{E}} R^e_{erm}(f)+\lambda \cdot \operatorname{penalty}\left(\left\{S^e(f)\right\}_{e \in \mathcal{E}}\right),
\end{equation}
where $R^e_{erm}(f) = \mathbb{E}_{p(x,y|e)}[\ell(f(x),y)]$ is the expected loss in environment $e$, and $S^e(f)$ denotes an environment-specific model statistic constrained by the penalty. 
Different invariant learning methods instantiate this statistic differently, such as the variance of environment risks in V-REx~\cite{v-rex}, calibration error in CLOvE~\cite{causal_ir}, or gradient-based optimality in IRMv1~\cite{IRM_training}. 
We build on IRMv1 because its gradient-based objective can regularize the encoder through environment-wise optimization signals, making it suitable for deep representation learning.
IRMv1 uses $S^e(f)=\left\|\nabla_w R^e_{erm}(w \circ \phi)\right\|^2$, where $w$ is a scalar dummy classifier fixed at 1.0. 
Although IRMv1 targets this goal, its bilevel structure and shared dummy classifier may cause unstable convergence, weak encoder supervision, and spurious overfitting~\cite{risks_ir}. These limitations cannot be fully eliminated, but can be mitigated in specific applications for more robust IRM~\cite{irm_eval}, motivating Section~\ref{invariant_alignment}.
}

\subsubsection{Split Environments for Invariant Learning}
Invariant learning relies on segmenting environments to highlight differences across them~\cite{IR_intro}. Early methods assumed prior knowledge of environment labels, which is often unavailable in practice~\cite{bottleneck_ir, empirical_ir,environment_label,inaccessible_label_2}. Recent approaches focus on invariant learning without predefined labels. Creager et al.~\cite{EIC} proposes estimating environment labels via prior clustering before applying invariant learning (EIIL), while others explore segmentation strategies such as natural clustering using dataset-provided labels or unsupervised clustering~\cite{unshuffling}. Experiments show natural clustering outperforms unsupervised methods. Regardless of the approach, effective segmentation requires environments to expose unstable information that can be ignored during invariant learning~\cite{IR_intro}.


%% file: Tex/Motivation.tex
\section{Motivation}
\label{motivation}

\begin{figure}
    \centering
    \includegraphics[width=0.95\linewidth]{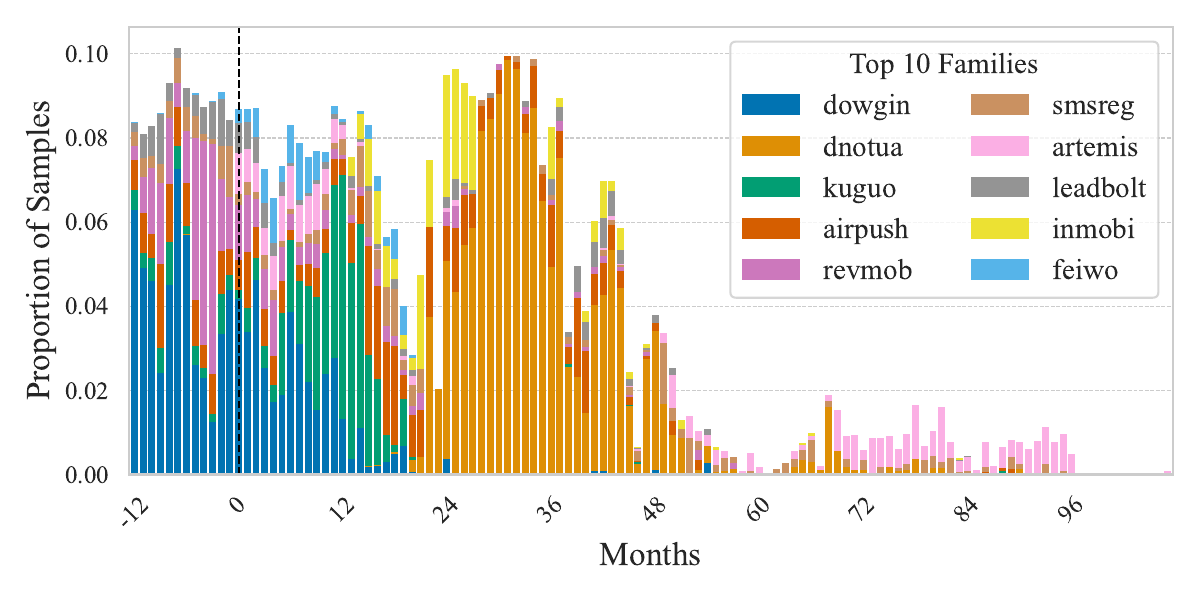}
    \caption{The top-10 families' proportions vary over time; zero marks the test start, negatives indicate training.}
    \label{fig:family_proportion}
\end{figure}

Learning-based malware detectors achieve strong performance in identifying threats; however, maintaining performance under drift remains challenging. 
Figure~\ref{fig:family_proportion} supports this claim, showing monthly proportion changes of the top-10 families in our dataset (Section~\ref{Dataset}) across training and testing phases. 
Some families dominant during training, such as Dowgin, gradually disappear in testing, while new families, such as Artemis, emerge. 
This evolution challenges detectors' cross-family generalization. 
Even for families like Airpush, present in both phases, feature distributions are affected by temporal proportion fluctuations and API changes.

\subsection{Threat Model}
\label{problem_scope}
Malware natural drift refers to the gradual evolution of malware distributions over time, driven by the emergence of new families and modifications to existing ones. Without knowledge of detection methods, attackers can alter malware features, causing distribution shifts that undermine the long-term effectiveness of detection systems.

Our threat model focuses on natural drift, which differs from adversarial attacks that rely on knowledge of the detection scheme to identify optimal perturbations for evading models~\cite{pierazzi2020problemspace}. We find that even natural drift disrupts the adaptation capabilities of traditional methods, hindering their ability to generalize to new distributions. Therefore, in this work, we address the long-term challenges posed by malware natural drift by enhancing the robustness of detection systems against malware evolution. We consider adversarial drift to be out of scope for this paper and plan to explore it in future work.



\begin{figure*}
    \centering
    \setlength{\abovecaptionskip}{0cm}
    \includegraphics[width=1.0\linewidth]{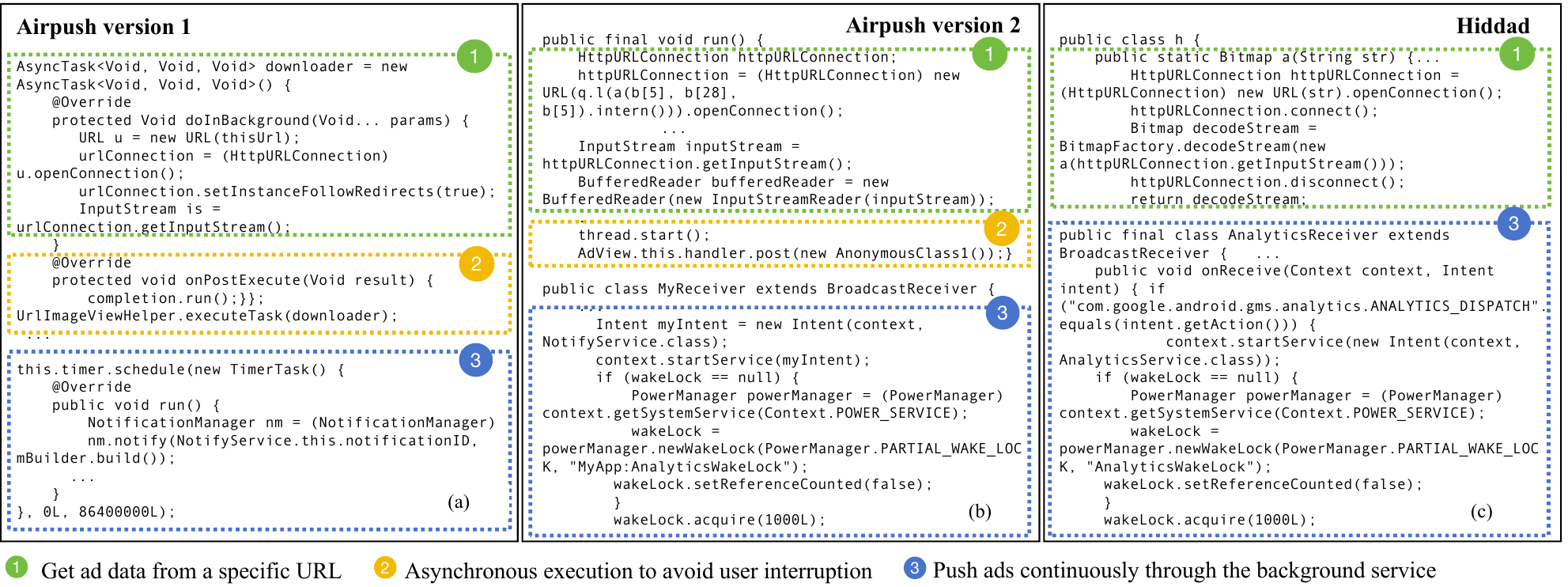}
    \caption{(a), (b), and (c) show real code snippets from an early Airpush version, a later Airpush version, and the Hiddad adware family. Airpush's core behavior includes: (1) getting ad data from a specific URL, (2) asynchronous execution to avoid user interruption, and (3) pushing ads continuously through the background service. (a) and (b) demonstrate that both Airpush versions share invariant behaviors, with similar API calls and permissions despite implementation differences. Hiddad, while skipping step (2) for simpler ad display, shares steps (1) and (3) with Airpush, especially the newer version.}
    \label{fig:motivation_sample}
\end{figure*}

\subsection{Invariance in Malware Evolution}
\label{sec:Malware Evolution}
\diff{The challenges highlighted in Section~\ref{problem_scope} motivate the search for training-set characteristics that generalize to future samples. 
Malware families and types provide a natural basis for such invariance, grouping malware by code structure, behavioral patterns, and malicious intent. 
Although implementations may vary due to evolution or obfuscation, variants within the same family often preserve behaviors consistent with their goals. 
Similarly, malware types capture broader operational intentions that may remain stable across families. 
These consistencies form two kinds of invariance: \textbf{intra-family invariance}, where variants within a family maintain consistent operational behaviors and attack strategies despite implementation differences; and \textbf{inter-family invariance}, where different families share common malicious patterns, such as resource abuse or evasion techniques.}

\diff{
To illustrate invariant behaviors under drift, we select APKs from Androzoo~\footnote{https://androzoo.uni.lu/} and decompile them with JADX\footnote{https://github.com/skylot/jadx}. 
We use Airpush variants across periods for intra-family invariance and Hiddad for inter-family invariance, as both families exhibit intrusive ad delivery despite different implementations. 
Figure~\ref{fig:motivation_sample} highlights invariant behaviors in code snippets: Airpush uses asynchronous task requests, while Hiddad relies on background services and scheduled tasks to evade detection.
}

Figure~\ref{fig:motivation_sample}(a)
and (b)
show core code from this family in 2014 and later years, respectively. The 2014 version uses \verb|NotifyService| and \verb|TimerTask| to notify users every 24 hours, maintaining ad exposure. The later version, adapting to Android 8.0’s restrictions, triggers \verb|NotifyService| via \verb|BroadcastReceiver| with \verb|WAKE_LOCK| to sustain background activity. In Drebin’s~\cite{Arpdrebin} feature space, these invariant behaviors are captured through features like \verb|android_app_NotificationManager;notify|, \verb|permission_READ_PHONE_STATE| and so on. Both implementations also use \verb|HttpURLConnection| for remote communication, asynchronously downloading ads and tracking user activity, and sharing Drebin features such as \verb|java/net/HttpURLConnection| and \verb|android_permission_INTERNET|.

Similarly, Figure~\ref{fig:motivation_sample}(c)
shows a real sample from the Hiddad family, which uses HTTP connections for ad delivery, along with \verb|AnalyticsService| and \verb|WAKE_LOCK| for continuous background services. Permissions like \verb|android_permission_WAKE_LOCK| and API calls such as \verb|getSystemService| reflect shared, cross-family invariant behaviors, whose learning would enhance model detection across variants.

Capturing the core malicious behaviors of Airpush aids in detecting both new Airpush variants and the Hiddad family, as they share similar malicious intents. These stable behaviors form consistent indicators in the feature space. However, detectors with high validation performance often fail to adapt to such variants, underscoring the need to investigate root causes and develop a drift-robust malware detector.

\begin{center}
\fcolorbox{black}{gray!10}{\parbox{.9\linewidth}{\textit{\textbf{Take Away}: Training features contain learnable invariance within and across malware families.}}}
\end{center}

\subsection{Failure of Learning Invariance}
\label{sec:invariance definition}
Let $f_r \in \mathcal{R}$ be a sample in the data space with label $y \in \mathcal{Y} = \{0, 1\}$, where 0 represents benign software and 1 represents malware. The input feature vector $x \in \mathcal{X}$ includes features $\mathcal{F}$ extracted from $f_r$ according to predefined rules. The goal of learning-based malware detection is to train a model $\mathcal{M}$ based on $\mathcal{F}$, mapping these features into a latent space $\mathcal{H}$ and passing them to a classifier for prediction. The process is formally described as follows:
\begin{equation}
\arg \min _{\theta} R_{erm}\left(\mathcal{F}\right),
\end{equation}
where $\theta$ is the model parameter and $R_{erm}(\mathcal{F})$ represents the expected loss based on features space $\mathcal{F}$, defined as:
\begin{equation}
R_{erm}\left(\mathcal{F}\right)=\mathbb{E}[\ell(\hat{y}, y)].
\end{equation}
$\ell$ is a loss function. By minimizing the loss function, $\mathcal{M}$ achieves the lowest overall malware detection error.

\subsubsection{Stability and Discriminability of Features}
\label{active ratio}
\diff{To investigate drift robustness in malware evolution from a feature perspective, we define stability and discriminability. 
Stability denotes a feature's consistent relevance across distributions, while discriminability reflects its ability to distinguish categories. 
To avoid model-induced biases, we provide a modelless definition applicable to diverse architectures.}

\diff{Let $f_j$ be the $j$-th feature in feature set $\mathcal{F}$, and let $S$ denote all samples. 
For any subset $S^{\prime} \subseteq S$, we define the active ratio of $f_j$ as the proportion of samples in which the feature is active. 
For binary features, where $f_j(s)=1$ indicates presence and $f_j(s)=0$ absence, the active ratio is:
\begin{equation}
\label{eq:active ratio}
r\left(f_j, S^{\prime}\right)=\frac{1}{\left|S^{\prime}\right|} \sum_{s \in S^{\prime}} f_j(s).
\end{equation}
We use this ratio to define feature stability and discriminability.}


\begin{myDef}
\textbf{Stable Feature}: A feature $f_j$ is stable if its active ratio remains consistent across temporal environments relative to the training distribution. 
Let the dataset be partitioned into a training set $S_0$ and temporal subsets $\mathcal{T}=\{S_{t_1}, S_{t_2}, \dots, S_{t_m}\}$ based on timestamps. 
Feature $f_j$ is stable if, for any sufficiently large temporal subset $S_{t_k} \in \mathcal{T}$:
\begin{equation}
\label{eq:stability}
\forall S_{t_k} \in \mathcal{T}, \left|S_{t_k}\right| \geq n_0, \quad \left|r\left(f_j, S_{t_k}\right)-r\left(f_j, S_0\right)\right| \leq \epsilon,
\end{equation}
where $\epsilon>0$ bounds deviation from the training distribution.
\end{myDef}

For discriminability, we consider sample categories. 
Let $C=\left\{C_1, C_2, \ldots, C_k\right\}$ be $k$ classes, and $S_k \subseteq S$ the samples in class $C_k$. 
The active ratio of $f_j$ in class $C_k$ is:
\begin{equation}
  r\left(f_j, S_k\right)=\frac{1}{\left|S_k\right|} \sum_{s \in S_k} f_j(s).  
\end{equation}

\diff{
\begin{myDef}
\textbf{Discriminative Feature}: A feature $f_j$ is discriminative if its active ratio differs significantly between two classes $C_p$ and $C_q$. 
Specifically, there exists $\delta > 0$ such that:
\begin{equation}
\label{eq:discrim}
  \exists C_p, C_q \in C, p \neq q, \quad\left|r\left(f_j, S_p\right)-r\left(f_j, S_q\right)\right| \geq \delta.
\end{equation}
\end{myDef}
Furthermore, feature discriminability should be independent of class proportions. 
For any subsets $\tilde{S}_p \subseteq S_p$ and $\tilde{S}_q \subseteq S_q$, where $\left|\tilde{S}_p\right| \neq\left|S_p\right|$ or $\left|\tilde{S}_q\right| \neq\left|S_q\right|$, the property holds:
\begin{equation}
    \left|r\left(f_j, \tilde{S}_p\right)-r\left(f_j, \tilde{S}_q\right)\right| \geq \delta.
\end{equation}}
\diff{\begin{myDef}
\textbf{Temporally Invariant Feature}: A feature $f_j$ is temporally invariant if it is stable relative to the training set $S_0$ and discriminative within each temporal environment $S_{t_k} \in \mathcal{T}$. 
Let $S_{t_k}^+$ and $S_{t_k}^-$ denote the malicious and benign subsets of $S_{t_k}$, respectively. 
Formally:
\begin{equation}
\label{eq:invariant}
\begin{split}
\forall S_{t_k} \in \mathcal{T}, \quad &
\underbrace{\left|r(f_j, S_{t_k}) - r(f_j, S_0)\right| \leq \epsilon}_{\text{Stability}} \\
\land \quad &
\underbrace{\left|r(f_j, S_{t_k}^+) - r(f_j, S_{t_k}^-)\right| \geq \delta}_{\text{Discriminability}}
\end{split}
\end{equation}
where $\epsilon$ is the stability bound, $\delta$ the discriminability threshold, and $r(\cdot)$ the active ratio defined in Eq.~\ref{eq:active ratio}.
\end{myDef}}

\begin{figure*}
    \centering
    \begin{subfigure}[t]{0.45\textwidth}  
        \centering
        \includegraphics[width=1.0\textwidth]{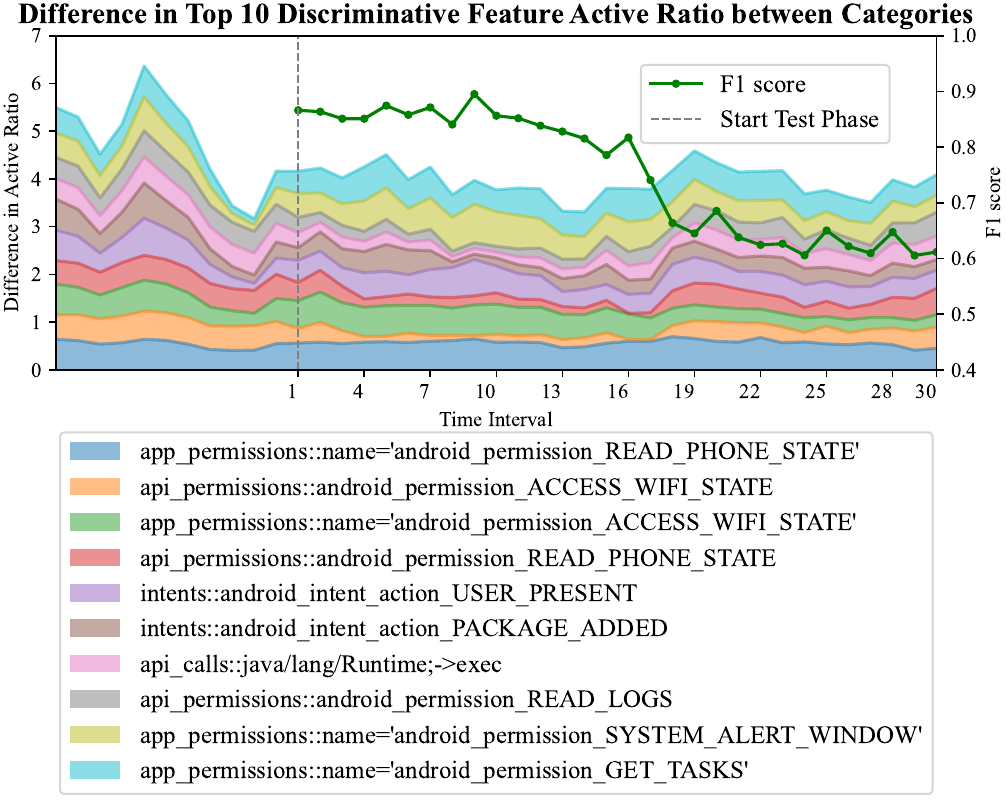}
        \caption{}
        \label{fig:diff}
    \end{subfigure}
    \hfill
    \begin{subfigure}[t]{0.45\textwidth}
        \centering
        \includegraphics[width=1.0\textwidth]{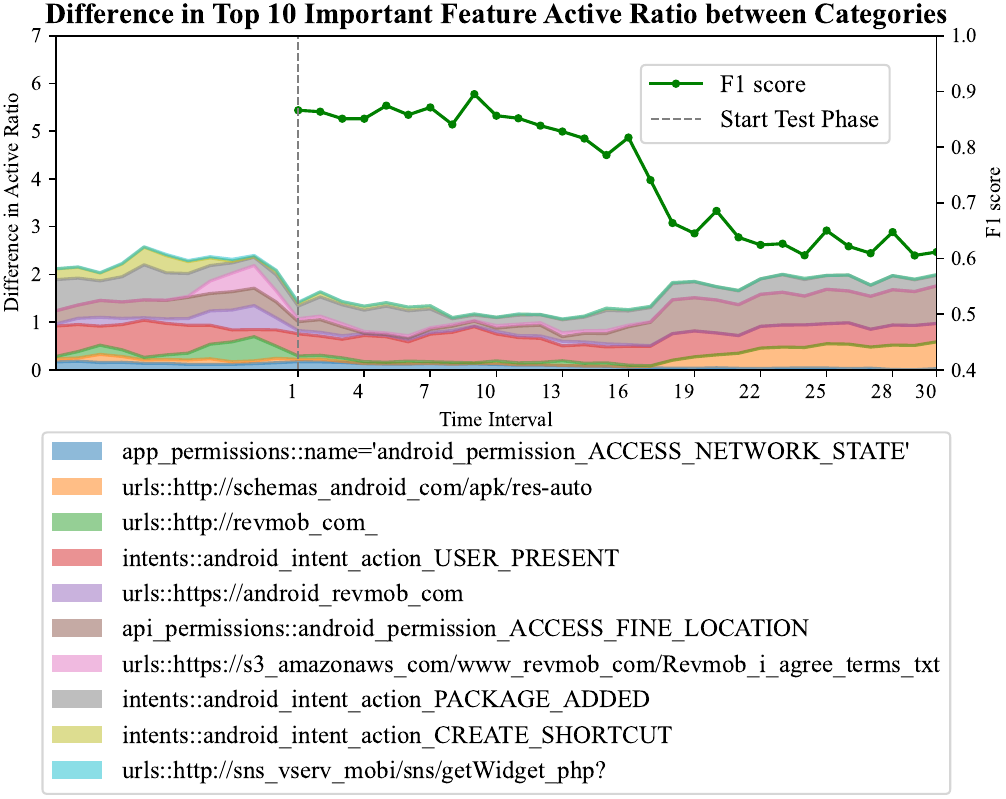}
        \caption{}
        \label{fig:importance}
    \end{subfigure}
    
    \caption{(a) and (b) illustrate changes in the discriminability of the top 10 discriminative training features and the top 10 important testing features, respectively. ``Discriminability'' is defined as the absolute difference in active ratios between benign and malicious samples. The grey dotted line indicates the start of the testing phase, with preceding values representing each feature's discriminability across months in the training set.}
    \label{fig:feature_discrimination}
\end{figure*}

\subsubsection{Failure Due to Learning Unstable Discriminative Features}
\label{motivation: failure}
\diff{The malware detector's strong performance within the same period indicates that detectors can learn discriminative features separating benign software from malware. 
However, performance degradation over time suggests that these features are not necessarily stable. 
To illustrate this, we sample 110,723 benign and 20,790 malware applications from Androzoo during 2014--2021, train on 2014 samples, and split the remaining data into 30 equal-sized temporal intervals. 
We extract Drebin~\cite{Arpdrebin} features, select the top 10 discriminative features by active-ratio difference, and track them over time. 
Using the DeepDrebin~\cite{Grossedeepdrebin} architecture, we evaluate each interval with F1 and estimate feature importance using Integrated Gradients with binary noise, averaged over five runs~\cite{IG_explain}. 
As shown in Figure~\ref{fig:feature_discrimination}, although the top discriminative features maintain stable active ratios, detector performance consistently declines.}


Figure~\ref{fig:feature_discrimination} compares active ratios of top discriminative features (a) and important features (b). 
Stable discriminative features persist, but ERM-based detectors often rely on unstable features with fluctuating effectiveness, causing poor drift generalization. 
Highly discriminative features, e.g., \verb|api_calls::java/lang/Runtime;->exec| and \verb|GET_TASKS|, are often linked to high-permission operations and potential malicious activity. 
These rarely appear in legitimate applications and reflect malware invariance, where core malicious intents persist despite evolving implementations.

\begin{center}
\fcolorbox{black}{gray!10}{\parbox{.9\linewidth}{\textit{\textbf{Take Away}: Training samples contain stable, highly discriminative invariant features, yet current malware detectors fail to learn them effectively.}}}
\end{center}

\subsection{Create Model to Learn Invariance}
\label{learn_invariant_feature}
\diff{This discussion highlights the importance of learning stable, discriminative features for drift-robust malware detection. 
ERM captures stable and unstable information correlated with the target variable~\cite{understanding}, often relying on unstable features when highly predictive. 
The key challenge is to isolate and enhance stable features, aligning with invariant learning principles in Section~\ref{invariant_learning}. 
Invariant learning faces two challenges in malware detection. 
First, effective environment segmentation is needed to reveal unstable information~\cite{environment_label, env_label}, but identifying variants that trigger shifts is uncertain. 
Second, high-quality encoder representations are essential for invariant predictors~\cite{yang2024invariant}, yet Figure~\ref{fig:feature_discrimination} shows that training features may fail to distinguish goodware from malware or capture malware execution intent, relying on ambiguous cues. 
To this end, we use temporal environment segmentation to expose instability in malware distribution drift. 
Within each environment, ERM guides target-related representation learning, while a shared encoder captures stable and unstable information across environments. 
The detector then improves generalization by minimizing invariant risk to filter unstable elements.}

\begin{center}
\fcolorbox{black}{gray!10}{\parbox{.9\linewidth}{\textit{\textbf{Take Away}: Invariant learning requires training environments that expose unstable information and encoders capable of learning rich representations.}}}
\end{center}

%% file: Tex/Methodology.tex
\section{Methodology}

\subsection{Problem setting}
\label{sec:problem setting}
\diff{
We use uppercase and lowercase letters to denote matrices and vectors, respectively, and $|\mathcal{D}|$ denotes the number of elements in set $\mathcal{D}$. 
In Android malware detection, samples $x \in \mathbb{R}^{d}$ are divided into labeled training data $\mathcal{D}_{tr} = \{(x_i^{tr},y_i^{tr})\}_{i=1}^{|\mathcal{D}_{tr}|}$, where $y_{i}^{tr} \in \{0, 1\}$ denotes benign/malicious labels, and unlabeled test data $\mathcal{D}_{ts}$. 
Training data $\mathcal{D}_{tr}$ come from period $\mathcal{T}_{tr}$, with each sample $x_i^{tr}$ assigned timestamp $t_i^{tr}$. 
Test data $\mathcal{D}_{ts}$ comprise future samples $t_i^{ts} > \mathcal{T}_{tr}$ that appear progressively over time. 
We assume training and test data share the same label space and have unknown semantic similarities. 
The model is $\mathcal{M} = c \circ \phi$, where $c$ is the classifier and $\phi$ the feature encoder. 
This work aims to enhance $\phi$ to capture invariant semantic features that generalize to unseen test data.}


\subsection{Overall Architecture}
\diff{Section~\ref{motivation: failure} highlights the need to learn stable and discriminative features for drift robustness. 
We propose a detector-agnostic temporal invariant training method that extracts rich features and suppresses unstable factors through invariant risk minimization. 
It comprises two components: \textbf{Multi-Proxy Contrastive Learning (MPC)}, which uses dynamic proxies to encode multi-family malware semantics into compact, discriminative embeddings; and \textbf{Invariant Gradient Alignment (IGA)}, which aligns gradients across environments to help the encoder learn stable representations.}

\begin{figure*}[t]
    \centering
    \includegraphics[width=1.0\linewidth]{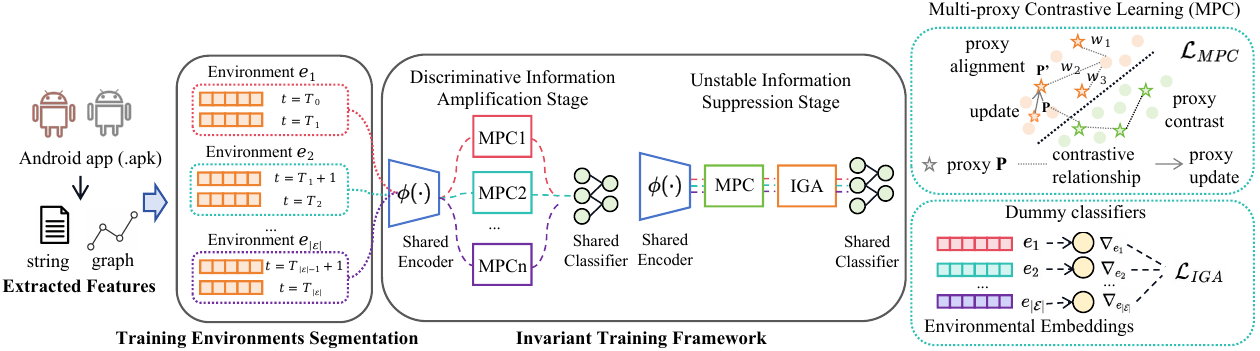}
    \caption{The proposed invariant training framework and its core components.}
    \label{fig:model_architecture}
\end{figure*}

Figure~\ref{fig:model_architecture} illustrates the overall framework of the proposed method, where Android applications are divided into non-overlapping environments based on their observation dates. The invariant training process includes two stages: 
\begin{itemize}
    \item Discriminative Information Amplification Stage: MPC is applied within each environment to minimize empirical risk, enabling the encoder to integrate discriminative features from each environment.
    \item Unstable Information Suppression Stage: Building on the model trained in the first stage, MPC aligns features across environments, followed by IGA fine-tuning with IRM to suppress unstable information.
\end{itemize}

\subsection{Training Environments Segmentation}
\label{environment_split}
\diff{Effective environment segmentation should expose unstable information. 
Since malware drift arises from multiple factors, such as market demand and Android version updates, we partition training data by application observation dates to capture their combined effects. 
For samples with timestamps from $T_{\min}$ to $T_{\max}$ and granularity $\Delta$, we divide the training data into $t$ time windows, each representing an environment $e \in \mathcal{E}$ with $|\mathcal{E}| = t$. 
The resulting ordered multiset is $\mathcal{D} = \{D_1,D_2,...,D_t\}$, where each $D_t$ contains $|D_t|$ samples. 
A sample $x_i$ with timestamp $t_i$ is assigned by:
\begin{equation}
\mathcal{E}(x_i)=\left\lfloor\frac{t_i-T_{\min }}{\Delta}\right\rfloor.
\end{equation}
As the environment index increases, timestamps approach the present. 
Time granularity affects representation learning by balancing distribution detail and sample sufficiency. 
Finer granularity risks sparse samples, while coarser granularity may obscure shifts. 
This balance requires considering label distribution to avoid misaligned representations. 
Supplementary materials explore the effects of granularity choices.}

\subsection{Multi-Proxy Contrastive Learning}
\label{multi-proxy contrastive learning}
\diff{Malware families have distinct distributions and imbalanced sample sizes, complicating category modeling in embedding space. 
Benign samples may also exhibit complex distributions due to factors such as geographic location or user behavior. 
Treating all samples within a family equally can exacerbate imbalance, while homogenizing them overlooks valuable variation. 
Thus, effective discrimination requires balancing intra-class compactness and diversity to obtain high-quality representations. 
To this end, we introduce multi-proxy contrastive learning, where each class is represented by multiple learnable proxies and samples are softly assigned according to behavioral or structural variations.}


For a batch of normalized feature embeddings $\mathbf{X}_c \in \mathbb{R}^{|\mathcal{B}_c| \times d}$ from category $c$ , where $|\mathcal{B}_c|$ is the batch size and $d$ is the feature dimension, we randomly initialize $K$ learnable proxies for both benign and malicious categories. Specifically, the proxies for class $c$ are denoted as $\mathbf{P}_c \in \mathbb{R}^{K \times d}$, where each row in $\mathbf{P}_c$ represents a proxy vector in the feature space. In this multi-proxy setup, each sample is softly assigned to all proxies in its category, which helps avoid over-reliance on any single proxy and mitigates the impact of noise. The similarity matrix between samples and all proxies in their category is computed and scaled by a temperature parameter $\tau$: 
\begin{equation}
\mathbf{S}_c = \mathbf{X}_c \mathbf{P}_c^\top/\tau, \quad \mathbf{S}_c \in \mathbb{R}^{|\mathcal{B}_c| \times K}.
\end{equation}
The assignment probability of the $i$-th sample is then calculated via a softmax:
\begin{equation}
p_{ij}^{(c)} = \exp(S_{c}^{(i, j)})/\sum_{k=1}^{K} \exp(S_{c}^{(i, k)}),
\end{equation}
where $S_{c}^{(i, j)}$ is the similarity between sample $i$ and proxy $j$ in class $c$. 
The alignment loss for class $c$ is:
\begin{equation}
\mathcal{L}_{pal}^{c} = -\frac{1}{|\mathcal{B}c|} \sum{i=1}^{|\mathcal{B}c|} \sum{j=1}^{K} p_{ij}^{(c)} \log p_{ij}^{(c)}.
\end{equation}
Finally, the total proxy alignment loss across all classes is:
\begin{equation}
\mathcal{L}_{pal} = \frac{1}{C} \sum_{c=1}^{C} \mathcal{L}_{pal}^{c}.
\end{equation}
Minimizing $\mathcal{L}_{pal}$ encourages samples of the same category to cluster around their most similar proxies, while still allowing flexibility in representation through multiple proxies. This not only enhances intra-class compactness and robustness to noise, but also captures potential sub-structures of each class.

\diff{The proxy distribution controls embedding diversity and compactness, so proxies should satisfy intra-class diversity and inter-class separation. 
We introduce two auxiliary regularization terms on proxy representations. 
\textbf{Intra-class diversity term}: 
This term encourages class proxies to spread across the hypersphere, enabling sub-cluster variation by maximizing their average pairwise distance:}
\begin{equation}
 \mathcal{L}_{intra}=-\frac{1}{C} \sum_{c=1}^C \frac{1}{\binom{K}{2}} \sum_{1 \leq i<j \leq K}\left\|\mathbf{P}_c^{(i)}-\mathbf{P}_c^{(j)}\right\|_2,
\end{equation}
where $\mathbf{P}_c^{(i)}$ is the $i$-th proxy in class $c$ and $K$ is the number of proxies per class. The negative sign encourages diversity by maximizing distances. \textbf{Inter-class separation term}: To promote discriminative power, $\mathcal{L}_{inter}$ enforces a minimum margin $m$ between class centers, preventing overlap or collapse of different category representations:
\begin{align}
\mathcal{L}_{inter} = \frac{1}{C(C-1)} 
\sum_{c_1 \neq c_2} \Big[ 
    & \left( m - \left\| \overline{\mathbf{P}}_{c_1} - \overline{\mathbf{P}}_{c_2} \right\|_2 \right) \notag \\
    & \cdot \mathbb{I}\left( \left\| \overline{\mathbf{P}}_{c_1} - \overline{\mathbf{P}}_{c_2} \right\|_2 < m \right)
\Big],
\end{align}
where $\bar{\mathbf{P}}_c = \frac{1}{K} \sum{j=1}^K \mathbf{P}_c^{(j)}$ denotes the center of class $c$. We combine these regularization terms with the proxy alignment term to obtain the total multi-proxy contrastive loss $\mathcal{L}_{MPC}$ in this module, which can be formalized as:
\begin{equation}
\mathcal{L}_{MPC} = \mathcal{L}_{pal} + \lambda_{intra} \cdot \mathcal{L}_{intra} + \lambda_{inter} \cdot \mathcal{L}_{inter},
\end{equation}
where $\lambda_{intra}$ and $\lambda_{inter}$ balance the weight of loss functions.

\subsection{Invariant Gradient Alignment}
\label{invariant_alignment}
\diff{Section~\ref{learn_invariant_feature} highlights two prerequisites for learning invariant features: environment segmentation that exposes unstable information and rich feature representations. 
Building on Invariant Risk Minimization (IRM), our goal is to guide the encoder toward stable representation aspects by aligning classifier gradients for samples of the same class across environments. 
IRMv1~\cite{IRM_training} targets this goal, but as discussed in Section~\ref{Learning Invariant Representation}, it can suffer from suboptimal performance due to complex bilevel optimization and a shared dummy classifier. 
To leverage IRM more effectively, we propose Invariant Gradient Alignment (IGA), which introduces environment-specific dummy classifiers, i.e., scalar parameters $s_e$ for each environment, to impose environment-aware gradient constraints. 
The objective function is shown in Eq.~\ref{grad_align}.}
\begin{equation}
\label{grad_align}
\mathcal{L}_{IGA} = \frac{1}{|\mathcal{E}|} \sum_{e \in \mathcal{E}} \left\|\nabla_{s_{e} \mid s_{e}=1.0} R^{e}(s_e \circ \phi)\right\|^2.
\end{equation}
Here, $s_{e}$ acts as an environment-specific scalar dummy classifier, initialized to 1.0 and updated through backpropagation. 
$\mathcal{L}_{IGA}$ evaluates how adjusting $s_e$ minimizes empirical risk in environment $e$. 
$R^e(s_e \circ \phi)$ is defined in Eq.~\ref{risk_minimization}:
\begin{equation}
\label{risk_minimization}
R^{e}(s_{e} \circ \phi)=\mathbb{E}^{e}\left[\mathcal{L}_{CLS} \left(s_{e}\left(\phi\right(x\left)\right), y\right)\right],
\end{equation}
where $\mathcal{L}_{CLS}^{e}$ denotes binary cross-entropy for malware detection. The term $\phi(x)$ represents the output of the shared encoder for samples $x, y \sim p(x, y|e)$ from environment $e$. The gradient penalty term encourages uniform feature representation by aligning gradients of classifiers of all environments, thereby promoting consistent model performance.

\subsection{Invariant Training Framework}
\label{invariant training}
Gradient adjustment for invariant learning classifiers relies heavily on the encoder’s ability to learn rich representations~\cite{rich}. Starting from random initialization often leads to suboptimal convergence. To overcome this, we propose a two-stage training strategy to first capture diverse features before applying invariant learning.

\subsubsection{Discriminative Information Amplification}
In the first stage, the multi-proxy contrastive learning module is applied to each environment to exploit discriminative features. 
This initialization amplifies the encoder's discriminative power. 
The optimization objective is defined in Eq.~\ref{erm loss}:
\begin{equation}
\label{erm loss}
\mathcal{L}_{DIA} = \frac{1}{|\mathcal{E}|} \sum_{e \in \mathcal{E}} (\mathcal{L}_{CLS}^e + \alpha \cdot \mathcal{L}_{MPC}^e),
\end{equation}
where $\mathcal{L}_{CLS}^e$ and $\mathcal{L}_{MPC}^e$ denote the classification and multi-proxy contrastive losses for environment $e$, respectively. 
Here, $\mathcal{L}_{CLS}^e$ is cross-entropy loss. 
Jointly minimizing empirical risk across environments equips the encoder with diverse feature representations for invariant training.


\subsubsection{Unstable Information Suppression}
\diff{To mitigate overfitting to environment-specific features, we reset the optimizer parameters before the second training phase. 
This reset helps the model refocus on invariant learning. 
We first apply multi-proxy contrastive loss across all samples to enhance class representations. 
Then, invariant gradient alignment harmonizes classification gradients across environments. 
The updated objective is defined in Eq.~\ref{irm loss}:}
\begin{equation}
\label{irm loss}
\mathcal{L}_{UIS} = \mathcal{L}_{CLS} + \alpha \cdot \mathcal{L}_{MPC} + \beta \cdot \mathcal{L}_{IGA},
\end{equation}
The hyperparameters $\alpha$ and $\beta$ balance the loss terms. 
This two-stage approach first captures broad discriminative features and then refines them for cross-environment invariance, improving generalization under distribution shifts. 
The complete invariant training algorithm is provided in the supplementary material. 



%% file: Tex/Experiment.tex
\section{Evaluation}
\label{evaluation}
\diff{This section evaluates our method's effectiveness in improving drift robustness across Android malware feature spaces. 
We also analyze component contributions to the invariant learning framework. 
The evaluation addresses four questions:
\begin{description}
    \item[RQ1.] Can our framework mitigate detector aging across feature spaces?
    \item[RQ2.] Can our framework stabilize detectors under different drift scenarios?
    \item[RQ3.] Does our framework learn invariant features?
    \item[RQ4.] Can our framework support continual learning?
\end{description}
To ensure reliability, we average results over three random seeds (1, 42, 2024). 
All experiments ran on an RTX A6000. 
The dataset and code will be released upon acceptance.}

\input{Table/RQ1}

\subsection{Evaluation Settings}
\subsubsection{Dataset}
\label{Dataset}
\diff{
To evaluate long-term robustness under malware evolution, we construct a dataset covering 2014 to June 2025 by extending the Transcendent~\cite{transcending, transcend} dataset with newly collected Androzoo samples from 2019 onward. We stop at June 2025 because newly released malware samples in 2025 are sparse on Androzoo, with fewer than 15 new malware samples per month on average. After filtering failed downloads and feature extraction errors, the dataset contains 308,637 benign apps and 48,707 malicious apps. Malware labels are derived from VirusTotal reports: following prior work, apps flagged by more than four security vendors (\(vt > 4\)) are labeled as malware. 
This relatively low threshold includes borderline cases, creating a challenging setting for evaluating detector robustness under noisy and evolving malware distributions. 
We use Euphony~\cite{euphony} to assign malware family labels, with unlabeled samples treated as unknown. Following TESSERACT~\cite{tesseract}, we mitigate temporal and spatial biases by training only on 2014 samples and testing on future samples from 2015 to June 2025. The 2014 training data are further split into 80\% proper training and 20\% validation sets. Detailed yearly statistics are provided in Appendix~\ref{app:dataset-statistics}.
} 

\diff{
For evaluation under temporal distribution shift, we organize normalized test data by time and family availability. 
We consider two settings: \textbf{Closed-world}, where test samples contain only malware families observed in training to test intra-family invariance, and \textbf{Open-world}, where test samples contain only families absent from training to test inter-family invariance. 
Because malware families appear imbalanced and non-uniformly over time, year-by-year evaluation may bias the comparison. 
Thus, for each setting, we sort test samples chronologically and partition them into 10 equal-sized temporal segments. 
These segments, indexed as 1-10 in tables and figures, represent successive intervals with comparable sample sizes rather than fixed calendar years.
}




\diff{\subsubsection{Candidate Detectors}
We evaluate \model\ across three representative Android malware feature spaces: binary behavioral features, graph-based features, and permission-language features. 
\textbf{Drebin}~\cite{Arpdrebin} represents APKs as binary feature vectors extracted from nine behavioral data types, such as permissions, API calls, and hardware components. 
Since \model\ is designed for neural architectures, we use \textbf{DeepDrebin}~\cite{Grossedeepdrebin}, which adopts Drebin's feature space with a three-layer DNN encoder and classifier. 
\textbf{Malscan}~\cite{malscan} uses graph-based features by extracting sensitive API calls and summarizing them with centrality measures. 
\textbf{BERTroid}~\cite{bertroid} encodes Android permission strings with a BERT-based model to capture contextual relationships among permissions. 
For fair comparison, we apply the same two-layer classifier to each feature representation, with the final layer performing binary classification.}






\diff{\subsubsection{Baseline}
We compare \model\ with representative static robustness methods and a continual learning baseline. 
For static baselines, \textbf{APIGraph}~\cite{apigraph} stabilizes API-based feature spaces by clustering semantically similar APIs, \textbf{T-stability}~\cite{svm_ce} constrains features with unstable temporal behavior, and \textbf{Guided Retraining}~\cite{guide_retraining} improves representations for difficult samples through supervised contrastive learning. 
Together, these baselines cover feature-level and representation-level robustness strategies under offline training. 
For continual learning, we use \textbf{HCC}~\cite{continuous}, which periodically detects drifted samples and updates the detector during deployment. 
Detailed implementation settings are provided in the supplementary material.}

\subsubsection{Metrics}
\diff{We use macro-F1 and Area Under Time (AUT). 
Macro-F1 handles class imbalance by balancing precision and recall on a fixed-time test set. 
AUT, proposed by TESSERACT~\cite{tesseract}, measures temporal performance:
\begin{equation}
\operatorname{AUT}(m, N)=\frac{1}{N-1} \sum_{t=1}^{N-1}\left(\frac{m(x_t+1)+m(x_t)}{2}\right),
\end{equation}
where $m$ is the performance metric, instantiated as F1-score, $m(x_t)$ denotes performance at time $t$, and $N$ is the number of monthly test slots. 
AUT ranges from $[0, 1]$, with 1 indicating perfect performance across all windows.}

\subsection{Enhance Different Feature Space (RQ1)}
\label{rq1}
\diff{
This section evaluates whether \model\ mitigates detector degradation across feature spaces. 
All classifiers are trained on 2014 samples and tested monthly from 2015 to 2025, with annual AUT(F1, 12m) results summarized in Table~\ref{tab:rq1}. 
We use monthly environment partitioning and discuss its granularity in supplementary materials. 
We compare against applicable baselines for each detector: T-stability~\cite{svm_ce} for Drebin~\cite{Arpdrebin}, APIGraph~\cite{apigraph} for Drebin and DeepDrebin, and Guided Retraining~\cite{guide_retraining} for all detectors. 
Overall, \model\ consistently delivers the strongest long-term performance across feature spaces.
}

\diff{
As drift increases over the test years, AUT declines for all methods, but \model\ achieves up to an 8\% improvement in the first year, reflecting practical scenarios where detectors are often retrained annually~\cite{apigraph}. 
Although AUT gains diminish over time, periodic retraining keeps performance degradation manageable. 
We further compare \model\ with a monthly retraining upper bound in supplementary materials, showing that \model\ closely approaches this bound during the first deployment year without requiring additional labels.
}

\diff{
T-stability performs conservatively on early test samples but degrades more slowly on distant ones, consistent with its goal of suppressing temporally unstable features. 
However, its feature-level constraint prioritizes stability over discriminability and does not strengthen stable discriminative feature combinations, limiting its overall performance compared to \model. 
Guided Retraining (GR) improves locally ambiguous representations near the decision boundary through error-conditioned contrastive learning, so its effectiveness depends on the number and structure of difficult samples produced by the base detector. 
This explains its mixed behavior across feature spaces: it may underperform when difficult samples are scarce or imbalanced, but improve when drifted samples resemble training-time difficult cases. 
Nevertheless, GR remains ERM-based, and its robustness is bounded by the generalization of training-time difficult samples, whereas \model\ explicitly promotes temporal invariance.
}

\begin{center}
\fcolorbox{black}{gray!10}{\parbox{.9\linewidth}{\textit{\textbf{Take Away}: \model\ slows down the aging of detectors in each feature space and has significant performance gains in the years just after deployment.}}}
\end{center}

\begin{figure}
    \centering
    \includegraphics[width=0.85\linewidth]{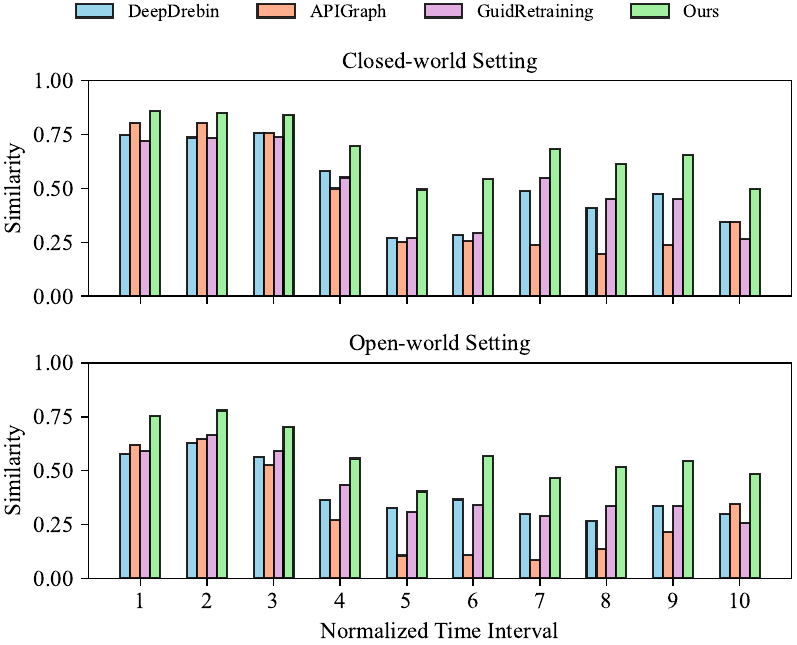}
    \caption{Cosine similarity of malware representations under closed-world and open-world settings, computed with DeepDrebin~\cite{Grossedeepdrebin} on Drebin features~\cite{Arpdrebin}. Higher values indicate more stable representations.}
    \label{fig:rq2}
\end{figure}
\input{Table/RQ2_2}

\subsection{Robustness in Different Drift Scenarios (RQ2)}
\label{rq2}
This section evaluates whether \model\ learns stable representations under different distribution drift.
We first evaluate the classification performance of the DeepDrebin~\cite{Grossedeepdrebin} detector against baselines and \model\ in different scenarios (closed-world and open-world), and compute AUT(F1, 12m) starting from the validation set.
Table~\ref{tab:rq2_2} shows that \model\ improves AUT(F1, 12m) by 5.78\% in the closed-world setting and by 3.17\% in the open-world setting, indicating its effectiveness in learning inter- and intra-invariant representations.

\diff{
To assess representation stability, we compare malware embeddings over time. 
For each temporal test interval, we average malware hidden representations and compute cosine similarity with the mean training malware embedding. 
Figure~\ref{fig:rq2} plots this similarity under closed-world and open-world settings, where higher values indicate greater temporal stability. 
\model\ consistently achieves higher similarity than all baselines in both settings.
}

\begin{center}
\fcolorbox{black}{gray!10}{\parbox{.9\linewidth}{\textit{\textbf{Take Away}: \model\ generates stable feature representations for both evolved variants of existing and new families.}}}
\end{center}

\subsection{Effective Invariant Feature Learning (RQ3)}
\label{rq3}
\diff{
We examine whether \model\ focuses on discriminative features that remain effective under temporal drift. 
As discussed in Section~\ref{motivation: failure}, ERM-based models often rely on transient correlations that degrade as distributions evolve. 
We use the Drebin feature space because its features are separable and interpretable. 
To quantify this behavior, we define the \emph{Feature Contribution Score} (FCS) for each feature $f_j$ as:
\begin{equation}
    FCS_j = \left|r\left(f_j, S_m\right) - r\left(f_j, S_b\right)\right| \cdot IS_j,
\end{equation}
where $r(f_j, S_m)$ and $r(f_j, S_b)$ are the activation ratios of $f_j$ in malware and benign samples, respectively, and $IS_j$ is its Integrated Gradients importance score for the malware class. 
The first term measures feature discriminability in the current evaluation set, while $IS_j$ measures model reliance. 
For a model $\mathcal{M}$, we sum FCS over all features to measure its overall emphasis on discriminative features.
}

\diff{
FCS itself does not encode time; temporal stability is reflected by whether high aggregate FCS persists across test segments. 
Transient features may have high FCS in one segment but decay as distributions shift, whereas invariant learning should slow this degradation. 
Using the closed-world and open-world protocols in Section~\ref{Dataset}, we compute FCS during training and across 10 temporal test segments. 
Invariant training yields higher and more persistent aggregate FCS than ERM-based baselines in both settings, suggesting that \model\ better maintains discriminative feature usage under temporal drift and complementing the representation-level analysis in RQ2. 
Detailed FCS values are in supplementary materials.
}

\diff{
We further conduct a qualitative case study to inspect the top features learned by \model. 
The results show that \model\ emphasizes intent-level behavioral prerequisites, such as broadcasts triggered by user activity or package installation and network availability checks. 
In contrast, ERM-based DeepDrebin focuses more on implementation-specific APIs and narrowly scoped permissions. 
The full feature rankings and case study are provided in the supplementary material.
}

\begin{center}
\fcolorbox{black}{gray!10}{\parbox{.9\linewidth}{\textit{\textbf{Take Away}: \model\ enables the detector to learn stable, discriminative features from the training set.}}}
\end{center}


\begin{figure}
    \centering
    \includegraphics[width=0.8\linewidth]{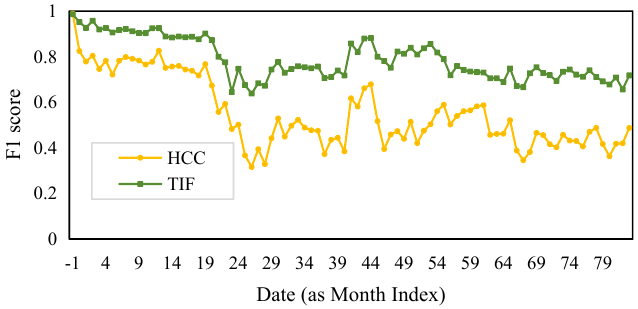}
    \caption{F1 scores for \model~and HCC~\cite{continuous} in the no continual learning scenario. A subscript of -1 indicates the result of the model on the initial validation set.}
    \label{fig:represent}
\end{figure}

\begin{figure}
    \centering
    \includegraphics[width=0.8\linewidth]{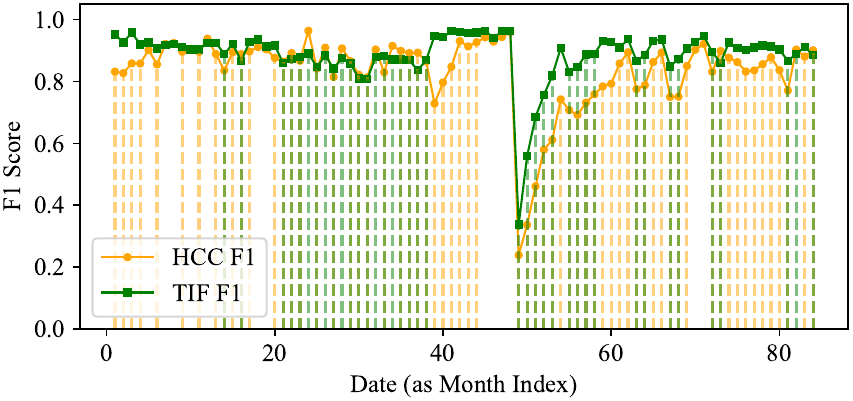}
    \caption{Detection performance of HCC~\cite{continuous} and the proposed \model-equipped framework on Android malware. Yellow dashed lines indicate update points of HCC; green lines represent \model. Overlapping updates appear as brown due to color blending.}
    \label{fig:active_learning}
\end{figure}

\subsection{Effectiveness on Continual Learning (RQ4)}
\label{sec:continual}
\diff{As discussed in Section~\ref{Drift-robust Malware Detectors}, continual learning mitigates detector aging by updating models with drifted samples over time. 
Since \model\ is model- and feature-agnostic, it can be integrated into continual learning frameworks to improve representation stability and reduce update cost. 
We evaluate this using HCC~\cite{continuous} under two complementary settings. 
The \emph{without-update} setting removes the update stage and isolates representation quality, while the \emph{with-update} setting evaluates the full drift-handling pipeline.}


\diff{
\subsubsection{Without Continual Updates}
As noted in Section~\ref{learn_invariant_feature}, high-quality representations of complex malware families are essential for continual learning. 
To isolate representation performance, we remove the update stage and compare \model\ with HCC at the representation level (Figure~\ref{fig:represent}). 
\model\ achieves an average F1 improvement of 49.16\% on drift samples over HCC, whose hierarchical contrastive learning is limited by numerous families and severe class imbalance. 
In contrast, \model\ balances representation granularity and sample size with multiple proxies, while invariant learning promotes stable features. 
These representations help the detector remain robust early in evaluation, delaying the first model update.
}

\diff{
\subsubsection{With Continual Updates}
For fair comparison with HCC in a realistic drift-handling scenario, we follow its update scheme but replace fixed monthly updates with an F1-based trigger in this controlled study. 
Specifically, the model is evaluated on the next month's labeled data; if F1 falls below 0.90, an update is triggered by adding a fixed budget of low-confidence samples to the training set. 
We use F1 only as an oracle-like trigger for analysis, not as a deployable update rule, because true F1 is unavailable without labels in real deployments. 
This design isolates representation lifespan under an idealized drift detector. 
In practice, the trigger can be replaced by uncertainty-based criteria~\cite{cade} or other drift detection conditions~\cite{transcend, transcending} without changing the framework.
} 

\diff{
Following Section~\ref{Dataset}, we perform a longitudinal evaluation (2015-2021) using Drebin features~\cite{Arpdrebin} trained on 2014 data. This period is selected to rigorously assess representation robustness against behavioral drift while minimizing the confounding impact of feature depletion caused by post-2021 advanced packing. We adopt a probability-based update strategy, adding 100 low-confidence samples per step, consistent with HCC's best performance setup~\cite{continuous}. As shown in Figure~\ref{fig:active_learning}, \model\ demonstrates superior stability: it reduces update frequency from 65 (HCC) to 38, cutting labeling costs by 41.5\% without compromising detection performance.
}

\begin{center}
\fcolorbox{black}{gray!10}{\parbox{.9\linewidth}{\textit{\textbf{Take Away}: \model\ learns stable representations that delay degradation under drift and reduce update frequency, lowering labeling cost without sacrificing accuracy.}}}
\end{center}


\input{Table/Ablation}

\subsection{Ablation Study}

This section evaluates the robustness of each component through an ablation study. Five configurations are tested: the base model, +MPC1, +MPC1 +MPC2, +MPC1 +IGA, and +MPC1 +IGA +MPC2, where MPC1/MPC2 represent multi-proxy contrastive learning in two training stage, and IGA denotes invariant gradient alignment which is only used in the second stage. Table~\ref{tab: ablation} shows the AUT(F1, 12m) results for each setup with the Drebin detector.

MPC1 improves intra-environment representations, surpassing the baseline. Adding MPC2 and IGA enhances generalization by aligning gradients and capturing global features. The full configuration achieves the highest robustness by integrating stable and discriminative feature learning.

\section{Threats to Validity}
\label{sec:threat}
Several factors may affect our results. 
First, \model\ introduces additional training components and hyperparameters; we mitigate this by evaluating multiple detectors and feature spaces under consistent protocols. 
Second, malware drift has complex and largely unobservable causes, so temporal partitioning serves as an agnostic proxy for observable evolution rather than decomposing drift factors. 
Third, our evaluation focuses on natural temporal drift in real-world Android malware and excludes adversarial shifts or adaptive attacks. 
Extended discussion is provided in the supplementary material.

\section{Supplementary Explanation}
The supplementary material provides additional details: (1) methodology, including the complete invariant training algorithm; 
(2) experimental setup, including implementation settings and dataset statistics; 
(3) extended results, including comparisons with ERM plus regularizers, which are standard ways to mitigate overfitting and improve generalization~\cite{regularization,regularization_understanding}, monthly post-deployment performance, and detailed FCS values; 
and (4) supporting analysis, including case studies, training overhead, and extended threats to validity.

%% file: Table/RQ1.tex
\begin{table*}[!t]
\caption{AUT(F1, 12m) of candidate detectors on future-year samples across feature spaces. 
All classifiers are trained on 2014 samples; w/o denotes the detector without robustness enhancement. 
$\Delta$ indicates relative improvement (\%) over the base detector. 
AG, GR, and TS denote APIGraph~\cite{apigraph}, Guided Retraining~\cite{guide_retraining}, and T-stability~\cite{svm_ce}. 
Since Drebin~\cite{Arpdrebin} and DeepDrebin~\cite{Grossedeepdrebin} share the same feature space, \model\ is applied once. 
Standard deviations are computed over three seeds.}
\centering
\label{tab:rq1}
\resizebox{0.9\linewidth}{!}{
\begin{tabular}{ccccccccccccc} 
\toprule
\multirow{2}{*}{\begin{tabular}[c]{@{}c@{}}Test\\Years\end{tabular}}                                      & \multicolumn{2}{c}{Drebin~\cite{Arpdrebin}}                                                              & \multicolumn{4}{c}{DeepDrebin~\cite{Grossedeepdrebin}}                                                                                                                                                                                                                      & \multicolumn{3}{c}{Malscan~\cite{malscan}}                                                                                                                                    & \multicolumn{3}{c}{BERTroid~\cite{bertroid}}                                                                                                                                     \\ 
\cmidrule(lr){2-13}& w/o   & TS    & w/o   & AG                                                                          & GR    & Ours                                                                 & w/o   & GR                        & Ours                                                                 & w/o   & GR   & Ours    \\ 
\midrule
\begin{tabular}[c]{@{}c@{}}2015\\$\Delta$\end{tabular}                                                   & 0.858 & \begin{tabular}[c]{@{}c@{}} 0.824\\\textcolor{red}{$\downarrow$} 3.96\end{tabular} & 0.859 & \begin{tabular}[c]{@{}c@{}}0.868\\ {\color{green} {\color{green} $\uparrow$}} 1.05\end{tabular}              & \begin{tabular}[c]{@{}c@{}}0.814\\\textcolor{red}{$\downarrow$} 5.24\end{tabular}  & \begin{tabular}[c]{@{}c@{}}$0.928_{\pm 0.004}$\\ {\color{green} {\color{green} $\uparrow$}} 8.03\end{tabular} & 0.842 & \begin{tabular}[c]{@{}c@{}}0.851\\{\color{green} $\uparrow$} 1.07\end{tabular}              & \begin{tabular}[c]{@{}c@{}}$0.902_{\pm 0.003}$\\ {\color{green} $\uparrow$} 7.13\end{tabular} & 0.771 & \begin{tabular}[c]{@{}c@{}}0.772\\{\color{green} $\uparrow$} 0.01\end{tabular}              & \begin{tabular}[c]{@{}c@{}}$0.833_{\pm 0.002}$\\ {\color{green} $\uparrow$} 8.04\end{tabular}  \\ 
\hline
\begin{tabular}[c]{@{}c@{}}2016\\$\Delta$\end{tabular} & 0.811 & \begin{tabular}[c]{@{}c@{}}0.782\\\textcolor{red}{$\downarrow$} 3.58\end{tabular} & 0.813 & \begin{tabular}[c]{@{}c@{}}0.818\\ {\color{green} $\uparrow$} 0.06\end{tabular}              & \begin{tabular}[c]{@{}c@{}}0.749\\\textcolor{red}{$\downarrow$} 7.87\end{tabular}  & \begin{tabular}[c]{@{}c@{}}$0.877_{\pm 0.004}$\\ {\color{green} $\uparrow$} 7.87\end{tabular} & 0.799 & \begin{tabular}[c]{@{}c@{}}0.795\\\textcolor{red}{$\downarrow$} 0.50\end{tabular} & \begin{tabular}[c]{@{}c@{}}$0.850_{\pm 0.005}$\\ {\color{green} $\uparrow$} 6.38\end{tabular} & 0.744 & \begin{tabular}[c]{@{}c@{}}0.743\\\textcolor{red}{$\downarrow$} 0.13\end{tabular} & \begin{tabular}[c]{@{}c@{}}$0.789_{\pm 0.004}$\\ {\color{green} $\uparrow$} 6.05\end{tabular}  \\ 
\hline
\begin{tabular}[c]{@{}c@{}}2017\\$\Delta$\end{tabular} & 0.750 & \begin{tabular}[c]{@{}c@{}}0.717\\\textcolor{red}{$\downarrow$} 4.40\end{tabular} & 0.752 & \begin{tabular}[c]{@{}c@{}}0.754\\ {\color{green} $\uparrow$} 0.03\end{tabular}              & \begin{tabular}[c]{@{}c@{}}0.656\\\textcolor{red}{$\downarrow$} 12.77\end{tabular} & \begin{tabular}[c]{@{}c@{}}$0.815_{\pm 0.006}$\\ {\color{green} $\uparrow$} 8.38\end{tabular} & 0.739 & \begin{tabular}[c]{@{}c@{}}0.734\\\textcolor{red}{$\downarrow$} 0.68\end{tabular} & \begin{tabular}[c]{@{}c@{}}$0.783_{\pm 0.006}$\\ {\color{green} $\uparrow$} 5.95\end{tabular} & 0.708 & \begin{tabular}[c]{@{}c@{}}0.713\\ {\color{green} $\uparrow$} 0.71\end{tabular}              & \begin{tabular}[c]{@{}c@{}}$0.752_{\pm 0.005}$\\ {\color{green} $\uparrow$} 6.21\end{tabular}  \\ 
\hline
\begin{tabular}[c]{@{}c@{}}2018\\$\Delta$\end{tabular} & 0.724 & \begin{tabular}[c]{@{}c@{}}0.718\\\textcolor{red}{$\downarrow$} 0.83\end{tabular} & 0.726 & \begin{tabular}[c]{@{}c@{}}0.726\\0.00\end{tabular}                               & \begin{tabular}[c]{@{}c@{}}0.623\\\textcolor{red}{$\downarrow$} 14.19\end{tabular} & \begin{tabular}[c]{@{}c@{}}$0.794_{\pm 0.006}$\\ {\color{green} $\uparrow$} 9.37\end{tabular} & 0.713 & \begin{tabular}[c]{@{}c@{}}0.708\\\textcolor{red}{$\downarrow$} 0.70\end{tabular} & \begin{tabular}[c]{@{}c@{}}$0.755_{\pm 0.006}$\\ {\color{green} $\uparrow$} 5.89\end{tabular} & 0.686 & \begin{tabular}[c]{@{}c@{}}0.694\\ {\color{green} $\uparrow$} 1.74\end{tabular}              & \begin{tabular}[c]{@{}c@{}}$0.726_{\pm 0.004}$\\ {\color{green} $\uparrow$} 2.19\end{tabular}  \\ 
\hline
\begin{tabular}[c]{@{}c@{}}2019\\$\Delta$\end{tabular} & 0.700 & \begin{tabular}[c]{@{}c@{}}0.726\\{\color{green} $\uparrow$} 3.71\end{tabular}              & 0.713 & \begin{tabular}[c]{@{}c@{}}0.713\\0.00\end{tabular}                               & \begin{tabular}[c]{@{}c@{}}0.575\\\textcolor{red}{$\downarrow$} 19.35\end{tabular} & \begin{tabular}[c]{@{}c@{}}$0.769_{\pm 0.007}$\\{\color{green} $\uparrow$} 7.85\end{tabular} & 0.700 & \begin{tabular}[c]{@{}c@{}}0.694\\\textcolor{red}{$\downarrow$} 0.86\end{tabular} & \begin{tabular}[c]{@{}c@{}}$0.741_{\pm 0.006}$\\{\color{green} $\uparrow$} 5.86\end{tabular} & 0.660 & \begin{tabular}[c]{@{}c@{}}0.670\\{\color{green} $\uparrow$} 2.42\end{tabular}              & \begin{tabular}[c]{@{}c@{}}$0.701_{\pm 0.006}$\\{\color{green} $\uparrow$} 6.36\end{tabular}  \\ 
\hline
\begin{tabular}[c]{@{}c@{}}2020\\$\Delta$\end{tabular} & 0.665 & \begin{tabular}[c]{@{}c@{}}0.723\\{\color{green} $\uparrow$} 8.72\end{tabular}              & 0.700 & \begin{tabular}[c]{@{}c@{}}0.696\\\textcolor{red}{$\downarrow$} 0.57\end{tabular} & \begin{tabular}[c]{@{}c@{}}0.527\\\textcolor{red}{$\downarrow$} 24.71\end{tabular} & \begin{tabular}[c]{@{}c@{}}$0.732_{\pm 0.008}$\\{\color{green} $\uparrow$} 4.57\end{tabular} & 0.687 & \begin{tabular}[c]{@{}c@{}}0.681\\\textcolor{red}{$\downarrow$} 0.87\end{tabular} & \begin{tabular}[c]{@{}c@{}}$0.726_{\pm 0.007}$\\{\color{green} $\uparrow$} 5.68\end{tabular} & 0.630 & \begin{tabular}[c]{@{}c@{}}0.639\\{\color{green} $\uparrow$} 2.70\end{tabular}              & \begin{tabular}[c]{@{}c@{}}$0.674_{\pm 0.007}$\\{\color{green} $\uparrow$} 7.14\end{tabular}  \\ 
\hline
\begin{tabular}[c]{@{}c@{}}2021\\$\Delta$\end{tabular} & 0.645 & \begin{tabular}[c]{@{}c@{}}0.716\\{\color{green} $\uparrow$} 11.01\end{tabular}             & 0.689 & \begin{tabular}[c]{@{}c@{}}0.686\\\textcolor{red}{$\downarrow$} 0.44\end{tabular} & \begin{tabular}[c]{@{}c@{}}0.497\\\textcolor{red}{$\downarrow$} 27.87\end{tabular} & \begin{tabular}[c]{@{}c@{}}$0.717_{\pm 0.009}$\\{\color{green} $\uparrow$} 4.06\end{tabular} & 0.676 & \begin{tabular}[c]{@{}c@{}}0.669\\\textcolor{red}{$\downarrow$} 1.04\end{tabular} & \begin{tabular}[c]{@{}c@{}}$0.713_{\pm 0.006}$\\{\color{green} $\uparrow$} 5.47\end{tabular} & 0.609 & \begin{tabular}[c]{@{}c@{}}0.623\\{\color{green} $\uparrow$} 3.28\end{tabular}              & \begin{tabular}[c]{@{}c@{}}$0.653_{\pm 0.006}$\\{\color{green} $\uparrow$} 7.72\end{tabular}  \\ 
\hline
\begin{tabular}[c]{@{}c@{}}2022\\$\Delta$\end{tabular} & 0.616 & \begin{tabular}[c]{@{}c@{}}0.687\\{\color{green} $\uparrow$} 11.53\end{tabular}             & 0.661 & \begin{tabular}[c]{@{}c@{}}0.658\\\textcolor{red}{$\downarrow$} 0.45\end{tabular} & \begin{tabular}[c]{@{}c@{}}0.450\\\textcolor{red}{$\downarrow$} 31.92\end{tabular} & \begin{tabular}[c]{@{}c@{}}$0.688_{\pm 0.011}$\\{\color{green} $\uparrow$} 4.84\end{tabular} & 0.643 & \begin{tabular}[c]{@{}c@{}}0.635\\\textcolor{red}{$\downarrow$} 1.24\end{tabular} & \begin{tabular}[c]{@{}c@{}}$0.680_{\pm 0.009}$\\{\color{green} $\uparrow$} 5.75\end{tabular} & 0.583 & \begin{tabular}[c]{@{}c@{}}0.597\\{\color{green} $\uparrow$} 3.43\end{tabular}              & \begin{tabular}[c]{@{}c@{}}$0.627_{\pm 0.007}$\\{\color{green} $\uparrow$} 8.06\end{tabular}  \\ 
\hline
\begin{tabular}[c]{@{}c@{}}2023\\$\Delta$\end{tabular} & 0.610 & \begin{tabular}[c]{@{}c@{}}0.662\\ {\color{green} $\uparrow$} 9.01\end{tabular}             & 0.638 & \begin{tabular}[c]{@{}c@{}}0.635\\\textcolor{red}{$\downarrow$} 0.47\end{tabular} & \begin{tabular}[c]{@{}c@{}}0.413\\\textcolor{red}{$\downarrow$} 34.64\end{tabular} & \begin{tabular}[c]{@{}c@{}}$0.665_{\pm 0.008}$\\ {\color{green} $\uparrow$} 4.23\end{tabular} & 0.615 & \begin{tabular}[c]{@{}c@{}}0.607\\\textcolor{red}{$\downarrow$} 1.30\end{tabular} & \begin{tabular}[c]{@{}c@{}}$0.652_{\pm 0.007}$\\ {\color{green} $\uparrow$} 6.02\end{tabular} & 0.558 & \begin{tabular}[c]{@{}c@{}}0.574\\ {\color{green} $\uparrow$} 2.87\end{tabular}              & \begin{tabular}[c]{@{}c@{}}$0.602_{\pm 0.008}$\\ {\color{green} $\uparrow$} 8.42\end{tabular}  \\
\hline
\begin{tabular}[c]{@{}c@{}}2024\\$\Delta$\end{tabular} & 0.601 & \begin{tabular}[c]{@{}c@{}}0.645\\ {\color{green} $\uparrow$} 7.32\end{tabular}             & 0.625 & \begin{tabular}[c]{@{}c@{}}0.623\\\textcolor{red}{$\downarrow$} 0.01\end{tabular} & \begin{tabular}[c]{@{}c@{}}0.421\\\textcolor{red}{$\downarrow$} 33.92\end{tabular} & \begin{tabular}[c]{@{}c@{}}$0.654_{\pm 0.009}$\\ {\color{green} $\uparrow$} 4.64\end{tabular} & 0.605 & \begin{tabular}[c]{@{}c@{}}0.595\\\textcolor{red}{$\downarrow$} 1.16\end{tabular} & \begin{tabular}[c]{@{}c@{}}$0.641_{\pm 0.009}$\\ {\color{green} $\uparrow$} 5.95\end{tabular} & 0.551 & \begin{tabular}[c]{@{}c@{}}0.565\\ {\color{green} $\uparrow$} 2.54\end{tabular}              & \begin{tabular}[c]{@{}c@{}}$0.593_{\pm 0.009}$\\ {\color{green} $\uparrow$} 7.62\end{tabular}  \\
\hline
\begin{tabular}[c]{@{}c@{}}2025\\$\Delta$\end{tabular} & 0.595 & \begin{tabular}[c]{@{}c@{}}0.636\\ {\color{green} $\uparrow$} 6.86\end{tabular}             & 0.616 & \begin{tabular}[c]{@{}c@{}}0.619\\{\color{green} $\uparrow$} 0.01\end{tabular} & \begin{tabular}[c]{@{}c@{}}0.416\\\textcolor{red}{$\downarrow$} 32.46\end{tabular} & \begin{tabular}[c]{@{}c@{}}$0.654_{\pm 0.008}$\\ {\color{green} $\uparrow$} 5.03\end{tabular} & 0.600 & \begin{tabular}[c]{@{}c@{}}0.589\\\textcolor{red}{$\downarrow$} 0.67\end{tabular} & \begin{tabular}[c]{@{}c@{}}$0.636_{\pm 0.010}$\\ {\color{green} $\uparrow$} 6.00\end{tabular} & 0.547 & \begin{tabular}[c]{@{}c@{}}0.559\\ {\color{green} $\uparrow$} 2.19\end{tabular}              & \begin{tabular}[c]{@{}c@{}}$0.588_{\pm 0.009}$\\ {\color{green} $\uparrow$} 8.04\end{tabular}  \\
\bottomrule
\end{tabular}
}
\end{table*}

%% file: Table/RQ2_2.tex
\begin{table}
\centering
\caption{AUT(F1, 12m) of the closed-world drift and open-world test scenarios after adding different schemes in the DeepDrebin~\cite{Grossedeepdrebin} detector. AG and GR represent baselines, APIGraph~\cite{apigraph} and Guide Retraining~\cite{guide_retraining}, respectively.}
\label{tab:rq2_2}
\renewcommand{\arraystretch}{1.1}
\setlength{\tabcolsep}{3pt}
\begin{tabular}{ccccccccc}                                         
\toprule
\multirow{2}{*}{\begin{tabular}[c]{@{}c@{}}Normalize \\Dataset No.\end{tabular}}  & \multicolumn{4}{c}{Closed world} & \multicolumn{4}{c}{Open world}  \\ 
\cmidrule(lr){2-9}
   & w/o   & AG    & GR    & Ours      & w/o   & AG    & GR    & Ours    \\ 
\midrule
1  & 0.911 & 0.920 & 0.906 & 0.949     & 0.903 & 0.905 & 0.871 & 0.918   \\
2  & 0.880 & 0.891 & 0.861 & 0.927     & 0.860 & 0.863 & 0.788 & 0.879   \\
3  & 0.855 & 0.865 & 0.808 & 0.902     & 0.817 & 0.822 & 0.714 & 0.839   \\
4  & 0.811 & 0.821 & 0.727 & 0.861     & 0.764 & 0.771 & 0.657 & 0.788   \\
5  & 0.777 & 0.789 & 0.663 & 0.827     & 0.727 & 0.735 & 0.602 & 0.752   \\
6  & 0.758 & 0.770 & 0.627 & 0.805     & 0.705 & 0.713 & 0.541 & 0.730   \\
7  & 0.742 & 0.754 & 0.598 & 0.788     & 0.689 & 0.698 & 0.497 & 0.715   \\
8  & 0.733 & 0.645 & 0.573 & 0.778     & 0.678 & 0.687 & 0.443 & 0.705   \\
9  & 0.728 & 0.739 & 0.552 & 0.771     & 0.671 & 0.679 & 0.692 & 0.697   \\
10 & 0.714 & 0.723 & 0.529 & 0.756     & 0.658 & 0.665 & 0.415 & 0.682   \\
\bottomrule
\end{tabular}
\end{table}

%% file: Table/Ablation.tex
\begin{table}
\centering
\caption{AUT(F1, 12m) after adding different components on DeepDrebin detector. $\bullet$ means with; $\circ$ means without.}
\renewcommand{\arraystretch}{1.1}
\setlength{\tabcolsep}{5pt}
\label{tab: ablation}
\begin{tabular}{cccccc} 
\toprule
MPC1 & $\circ$ & $\bullet$ & $\bullet$ & $\bullet$     & $\bullet$      \\
MPC2 & $\circ$ & $\circ$   & $\bullet$ & $\circ$       & $\bullet$      \\
IGA  & $\circ$ & $\circ$   & $\circ$   & $\bullet$     & $\bullet$      \\ 
\midrule
2015 & 0.859   & 0.902     & 0.913     & 0.900         & \uline{0.928}  \\
2016 & 0.813   & 0.847     & 0.854     & 0.856         & \uline{0.877}  \\
2017 & 0.752   & 0.783     & 0.782     & 0.793         & \uline{0.815}  \\
2018 & 0.726   & 0.760     & 0.750     & 0.771         & \uline{0.794}  \\
2019 & 0.713   & 0.747     & 0.742     & 0.765         & \uline{0.769}  \\
2020 & 0.700   & 0.719     & 0.731     & \uline{0.743} & 0.732          \\
2021 & 0.689   & 0.704     & 0.720     & \uline{0.731} & 0.717          \\
2022 & 0.661   & 0.667     & 0.686     & \uline{0.693} & 0.688          \\
2023 & 0.638   & 0.637     & 0.658     & 0.662         & \uline{0.665}  \\
\bottomrule
\end{tabular}
\end{table}

%% file: Tex/Conclusion.tex
\section{Conclusion}
Android malware detectors suffer performance degradation due to natural distribution changes caused by malware evolution. We identify learnable invariant patterns among malware samples with similar intent, enabling a drift-stable feature space. To address this, we propose a temporal invariant training framework that organizes samples chronologically and integrates multi-proxy contrastive learning with invariant gradient alignment. Experiments demonstrate improved robustness across feature spaces and drifting scenarios, promoting stable and discriminative representations.

%% file: Tex/Appendix.tex
\subsection{Malware Samples in Motivation Example}
\label{supp:motivation-samples}
The malware samples used in Figure~\ref{fig:motivation_sample} are: 

Figure~\ref{fig:motivation_sample}(a): Airpush variant, MD5: \textit{17950748f9d37bed2f660daa7a6e7439};

Figure~\ref{fig:motivation_sample}(b): Airpush variant, MD5: \textit{ccc833ad11c7c648d1ba4538fe5c0445};

Figure~\ref{fig:motivation_sample}(c): Hiddad sample, MD5: \textit{84573e568185e25c1916f8fc575a5222}

All APKs were obtained from Androzoo and decompiled using JADX.

\subsection{Invariant Training Algorithm}
\label{app:alg}
\input{Alg/Invariant_training}
This learning framework consists of two phases: discriminative information amplification and unstable information suppression. 
In Stage 1, the model learns discriminative representations within each temporal environment. 
Specifically, the classification loss $\mathcal{L}^{e}_{CLS}$ and the multi-proxy contrastive loss $\mathcal{L}^{e}_{MPC}$ are computed for each environment in Lines 6-7, and combined into the discriminative information amplification objective $\mathcal{L}_{DIA}$ in Line 9. 
The model is then updated by backpropagating $\mathcal{L}_{DIA}$ in Line 10. Stage 2 suppresses environment-specific unstable information through invariant training. 
The optimizer is reset in Line 14, while the model parameters learned from Stage 1 are retained in Line 15. 
The invariant gradient alignment loss $\mathcal{L}_{IGA}$, the classification loss $\mathcal{L}_{CLS}$, and the multi-proxy contrastive loss $\mathcal{L}_{MPC}$ are computed in Lines 21-23 and combined into the unstable information suppression objective $\mathcal{L}_{UIS}$ in Line 24. 
The model is then updated by backpropagating $\mathcal{L}_{UIS}$ in Line 25. 
This two-stage process first amplifies discriminative information and then suppresses unstable information, improving robustness and generalization under temporal distribution drift.

\subsection{Dataset Distribution}
\label{app:dataset-statistics}
\input{Table/Dataset_overall}

Table~\ref{tab: dataset} reports the yearly distribution of benign applications and malware grouped by family availability. 
Samples from 2014 are used for training; in the test years, malware families are categorized as seen, unseen, or unknown according to their Euphony labels.

\subsection{Extended Candidate Detectors and Baselines}
\label{app:detectors-baselines}

\subsubsection{Candidate Detectors}
In Android malware detection, an APK file contains executable code, such as \texttt{.dex} files, and configuration files, such as \texttt{manifest.xml}. 
These artifacts provide behavioral signals including API calls, permissions, and structural relations, which can be organized into different feature spaces. 
We evaluate \model\ on three representative feature spaces.

\textbf{Drebin and DeepDrebin.}
Drebin~\cite{Arpdrebin} represents Android applications using binary feature vectors extracted from nine data types, including hardware components, requested permissions, app components, intent filters, restricted API calls, used permissions, suspicious API calls, network addresses, and URLs. 
The original Drebin detector uses these features with a linear classifier. 
Since \model\ focuses on neural network architectures, we use DeepDrebin~\cite{Grossedeepdrebin}, which adopts the same feature space but employs a three-layer deep neural network for feature extraction and classification.

\textbf{Malscan.}
Malscan~\cite{malscan} adopts a graph-based feature space. 
It extracts sensitive API calls from APKs and represents them using four centrality measures: degree, Katz, proximity, and harmonic wave centralities. 
Following the original paper, we concatenate these centrality-based features for malware detection.

\textbf{BERTroid.}
BERTroid~\cite{bertroid} uses Android application permission strings as input and encodes them into dense representations with a BERT-based pre-trained model. 
This representation captures contextual relationships among permissions. 
In our experiments, we use the pooled BERT output as the feature representation for malware detection.

For fair evaluation, we apply a two-layer linear classifier to each feature representation, with the final layer performing binary classification.

\subsubsection{Static Baselines}
We select representative static robustness methods that improve malware detector robustness in a fixed offline training setting. 
These methods strengthen existing detectors without redefining the detection task or requiring continual updates during deployment. 
They cover different levels of abstraction, ranging from explicit feature manipulation to implicit representation enhancement.

\textbf{APIGraph.}
APIGraph~\cite{apigraph} represents an explicit feature selection strategy. 
It assumes that malware drift mainly arises from implementation-level API changes while semantic behaviors remain stable. 
APIGraph stabilizes the feature space by clustering semantically similar APIs. 
Following the original method, we derive 2,000 API clusters and replace all features within the same cluster with the corresponding cluster index.

\textbf{T-stability.}
T-stability~\cite{svm_ce} represents an explicit feature constraint strategy. 
It assumes that feature drift follows consistent temporal trends and constrains features with unstable temporal behavior. 
This method directly models feature-space drift under strong assumptions about feature evolution.

\textbf{Guided Retraining.}
Guided Retraining~\cite{guide_retraining} represents an implicit representation enhancement strategy. 
Instead of explicitly modeling drift sources, it identifies difficult samples through the confusion matrix and refines their representations using supervised contrastive learning. 
It also enables separate processing of easy and difficult samples during inference to improve detection accuracy.

Together, these methods span a spectrum from assumption-driven feature-level robustness to representation-level enhancement, enabling a principled comparison of static robustness strategies.

\subsubsection{Continual Learning Baseline}
To evaluate whether \model\ can reduce deployment cost in a continual learning setting, we compare it with \textbf{HCC}~\cite{continuous}, a state-of-the-art continual learning framework for Android malware detection. 
HCC addresses distribution drift by detecting drifted samples during deployment and periodically updating the detector with newly observed data. 
It combines enhanced representations with drift sample detection and model updating, making it a comprehensive continual learning baseline.

\begin{figure}[!t]
    \centering
    \begin{subfigure}{0.15\textwidth}
        \centering
        \includegraphics[width=0.9\linewidth]{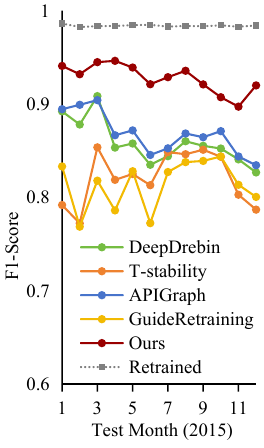}
        \caption{}
        \label{fig:sub1}
    \end{subfigure}%
    \hfill
    \begin{subfigure}{0.15\textwidth}
        \centering
        \includegraphics[width=0.9\linewidth]{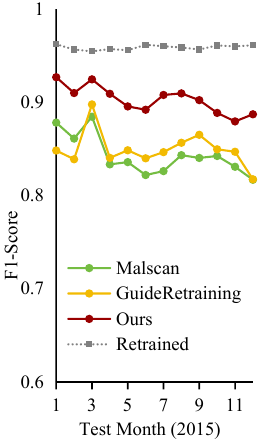}
        \caption{}
        \label{fig:sub2}
    \end{subfigure}%
    \hfill
    \begin{subfigure}{0.15\textwidth}
        \centering
        \includegraphics[width=0.9\linewidth]{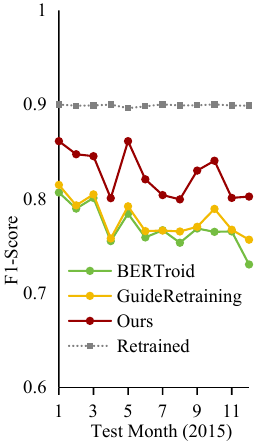}
        \caption{}
        \label{fig:sub3}
    \end{subfigure}
    \caption{Monthly performance of DeepDrebin (a), Malscan (b), and BERTroid (c) during the first test year (2015), with models initially trained on 2014 data. \textit{Retrained} denotes detectors updated monthly with labeled test samples.}
    \label{fig:rq1}
    \hfill
\end{figure}

\subsection{Monthly Performance in the First Deployment Year}
\label{supp:rq1-monthly}

Figure~\ref{fig:rq1} reports monthly F1 performance during the first test year, 2015, for DeepDrebin, Malscan, and BERTroid. 
All models are initially trained on 2014 data. 
The grey dashed line denotes a monthly retraining upper bound: for each month $n$, the training and validation sets are augmented with labeled samples up to month $n$, simulating an ideal setting where expert-labeled data are available for retraining. 
Compared with this upper bound, \model\ achieves consistently close monthly performance without using additional test-year labels. 
This indicates that \model\ improves early deployment robustness by learning more temporally stable representations rather than relying on frequent retraining.

\input{Table/RQ3}
\subsection{Detailed FCS Results}
\label{supp:fcs-case-study}

Table~\ref{tab:rq3} reports aggregate Feature Contribution Scores (FCS) during training and across the 10 temporal test segments under closed-world and open-world settings. 
The results show that \model\ maintains higher and slower-decaying FCS values than ERM-based baselines, indicating that invariant training preserves reliance on discriminative features under temporal drift.

\input{Table/case}
\subsection{Case Study}
\label{sec:case study}

We conduct a case study to illustrate that \model\ achieves robustness by capturing the invariant prerequisites of malware behavior, rather than overfitting to implementation-specific artifacts. We focus on two representative families, Airpush and Hiddad (Table~\ref{tab:case_feature}), both belonging to the adware category and align the motivation section (Section~\ref{motivation}). Although they differ in concrete implementations and evolution timelines, both families pursue the same high-level objective: aggressively delivering advertisements in response to user activity, which imposes a set of fundamental behavioral requirements independent of family or version.

Across both intra-family (Airpush$_{tr}-$Airpush$_{te}$) and inter-family (Airpush$_{tr}-$Hiddad$_{te}$) evaluations, \model\ consistently emphasizes features corresponding to these behavioral prerequisites. 
In particular, adware first detects when to activate, typically when the user becomes active or new applications are installed, and then determines how to deliver content by verifying network availability. 
Accordingly, \model\ assigns high importance to lifecycle-level system broadcasts (e.g., \texttt{USER\_PRESENT}, \texttt{PACKAGE\_ADDED}) and coarse-grained network checks (e.g., \texttt{ACCESS\_NETWORK\_STATE}). 
These signals abstract away from specific APIs or libraries and instead reflect the intent-driven logic shared by adware across families and over time.

In contrast, the ERM-based DeepDrebin~\cite{Grossedeepdrebin} baseline focuses on lower-level implementation artifacts that realize these behaviors in specific variants. For example, it prioritizes concrete API calls such as \texttt{getRunningTasks} or \texttt{requestLocationUpdates}, as well as narrowly scoped permissions like \texttt{ACCESS\_WIFI\_STATE} or \texttt{SYSTEM\_ALERT\_WINDOW}. 
These features implicitly assume specific implementation choices for realizing an intent (e.g., monitoring foreground activity via particular APIs or relying on Wi-Fi–specific connectivity checks). While such features may be predictive within a particular family or Android version, they encode assumptions about how a behavior is implemented rather than why it is necessary. As these assumptions change with OS evolution or family-specific design choices, the learned signals become brittle under temporal and natural shifts.

Overall, this contrast highlights that intent-level signals abstract away from concrete implementation choices, enabling \model\ to remain robust across family-specific variations and temporal evolution, whereas implementation-level features lead to brittle behavior under distribution shifts.

\subsection{Overhead Analysis}
\label{sec:overhead}
\input{Table/overhead}
Table~\ref{tab:overhead} reports the average per-epoch training time under a fixed batch size for each feature space. 
\model\ incurs higher training cost than standard ERM baselines (w/o), as expected from its design. 
Specifically, \model\ partitions each batch into temporal environments, optimizes additional objectives such as multi-proxy contrastive learning, and adopts a two-stage training procedure, increasing computation per iteration. 
This overhead occurs only during offline training and introduces no inference-time cost. 
Moreover, the increased training cost is offset by improved robustness under temporal drift. 
As shown in our continual learning evaluation (Section~\ref{sec:continual}), \model\ significantly reduces required model updates during deployment, lowering long-term retraining and labeling costs. 
For BERTroid, absolute training time is dominated by the BERT-based encoder, whose self-attention computation is substantially costlier than lightweight neural or linear models, making \model's relative overhead unavoidable but still practical. 
Consequently, robustness-enhancement methods introduce higher absolute training overhead in this setting, while relative overhead remains consistent with other feature spaces.

\input{Table/Env}
\subsection{Environment Granularity Selection}
\label{env seg}
Environment segmentation plays a critical role in invariant learning by controlling the trade-off between exposing temporal variation and ensuring sufficient samples per environment. We evaluate three segmentation strategies: monthly, quarterly, and equal-sized splits ($n=4, 8, 12$), with results reported in Table~\ref{tab: env}. Equal-sized splits often introduce severe label imbalance, weakening environment-wise invariance estimation. Temporal segmentation better preserves natural evolution patterns. Among them, finer-grained (monthly) segmentation is effective under mild, short-term drift by exposing subtle variations and leads to strong performance in the early deployment stage, which aligns with our focus on initial deployment robustness. However, overly fine-grained partitions may reduce per-environment representativeness and provide insufficient distributional contrast under long-term drift. In contrast, quarterly segmentation introduces stronger temporal contrast while maintaining adequate samples per environment, yielding better generalization across larger temporal gaps. Overall, these results indicate that the optimal granularity depends on the scale of temporal drift rather than a single universally optimal choice.

\input{Table/Regularization}

\subsection{Comparison to the Regularization Method}
ERM-trained models are easy to overfit to unstable features, limiting test set generalization~\cite{regularization}. Regularization methods, such as early stopping, $\ell$2 regularization, and dropout, help mitigate this by constraining model parameters~\cite{regularization_understanding}. Unlike regularization, invariant learning focuses on stable features under drift scenarios. We compare these regularization methods with our invariant learning framework during the whole test phase (Table~\ref{tab: regularization}). For reference, DeepDrebin (ERM) includes no regularization, while DeepDrebin~\cite{Grossedeepdrebin} employs dropout with a hyperparameter of 0.2. Table~\ref{tab: regularization} shows dropout improves performance under significant drift, outperforming early stopping and $\ell$2 regularization, which fails to capture optimal features. Invariant training consistently outperforms, capturing stable, discriminative features that maintain performance across distributions.

\subsection{Extended Threats to Validity}
\label{app:extended-threats}
We provide a more detailed discussion of the factors that may affect the interpretation and generalization of our results.

\textbf{Internal Validity.}
\model\ introduces additional components and a two-stage training procedure, which may raise concerns about implementation complexity and sensitivity to hyperparameter choices. 
To mitigate this risk, we evaluate \model\ across multiple malware detectors and feature spaces while keeping training protocols consistent with prior work. 
The observed improvements remain consistent across these settings, suggesting that the results are not tied to a particular model configuration. 
Nevertheless, different hyperparameter choices or implementation details may affect absolute performance, and further tuning could lead to different trade-offs between robustness and training cost.

\textbf{Construct Validity.}
Malware drift is a complex and multifaceted phenomenon whose underlying causes are largely unobservable. 
This work does not explicitly model specific sources of drift, such as API evolution, family emergence, or changes in app-market distributions. 
Instead, we adopt temporal partitioning as an agnostic proxy for the observable consequence of malware evolution, namely performance degradation over time. 
While this abstraction does not disentangle individual drift factors, it follows common practice in temporal robustness studies and enables a controlled and reproducible evaluation setting. 
Future work may combine temporal partitioning with more fine-grained drift attribution to better characterize different sources of instability.

\textbf{External Validity.}
Our evaluation focuses on natural temporal drift in real-world Android malware datasets. 
Therefore, the conclusions mainly apply to robustness under naturally occurring malware evolution. 
We do not consider adversarially crafted distribution shifts or adaptive attacks, where attackers deliberately manipulate samples with knowledge of the detector. 
Such adversarial drift represents an important but distinct research direction, and extending \model\ to this setting remains future work.

%% file: Alg/Invariant_training.tex
\begin{algorithm}[htb]
\caption{Algorithm of Invariant Training}
\label{alg1}
    \begin{algorithmic}[1]
    \REQUIRE ~training dataset containing $|\mathcal{E}|$ environments $\mathcal{D}_{tr}$ and each sample $x$ has a binary classification label $y_{c}, c \in [0, 1]$ and an environment label $y_e, e \in \mathcal{E}$, batch size $|\mathcal{B}|$, predefined number of epochs for stage 1 training $N$, hyperparameters $\alpha$ and $\beta$ for loss weighting, model $f$ with encoder network $\phi$ and predictor $h$. 
    \ENSURE 
    \STATE \textbf{Stage 1: Discriminative Information Amplification}
    \FOR{$\text{epoch} = 1$ to $N$}
        \FOR{$i = 1, \ldots, |\mathcal{B}|$}
            \FOR{$e \in \mathcal{E}$}
            \STATE select mini-batch of samples ${x_i^{e}} \subseteq \mathcal{B}$ \\
            \STATE calculate classification loss $\mathcal{L}^{e}_{CLS}$\\
            \STATE calculate multi-proxy contrastive loss $\mathcal{L}^{e}_{MPC}$ \\
            \ENDFOR
        \STATE obtain loss for empirical risk minimization $\mathcal{L}_{DIA} = \frac{1}{|\mathcal{E}|} \sum_{e \in \mathcal{E}} \mathcal{L}^{e}_{CLS} + \alpha \cdot \mathcal{L}^{e}_{MPC}$
        \STATE update $f$ by backpropagating $\mathcal{L}_{DIA}$
        \ENDFOR
    \ENDFOR
    \STATE \textbf{Stage 2: Unstable Information Suppression}
    \STATE Reset optimizer parameters
    \STATE Continue with model parameters from Stage 1\\
    \FOR{$\text{epoch} = N+1$ to $\text{TotalEpochs}$}
        \FOR{$i = 1, \ldots, |\mathcal{B}|$}
            \FOR{$e \in \mathcal{E}$}
            \STATE initialize dummy classifier $\{s_{e} = 1.0\}$ \\
            \ENDFOR
            \STATE calculate invariant gradient alignment loss $\mathcal{L}_{IGA}$ \\
            \STATE obtain classification loss $\mathcal{L}_{CLS}$ \\
            \STATE obtain multi-proxy contrastive loss $\mathcal{L}_{MPC}$ \\
            \STATE $\mathcal{L}_{UIS} = \mathcal{L}_{CLS} + \alpha \cdot \mathcal{L}_{MPC} + \beta \cdot \mathcal{L}_{IGA}$ \\
            \STATE update $f$ by backpropagating $\mathcal{L}_{UIS}$
        \ENDFOR
    \ENDFOR
    \end{algorithmic}
\end{algorithm}

%% file: Table/Dataset_overall.tex
\begin{table*}
\centering
\caption{Evaluation dataset. The dataset contains 357,344 applications from 2014 to 2025. The malware rate is about 10\% per year. 2014 is the training year; there is no unseen family. Notably, \textit{Unknown} is treated as one of \textit{Seen} families in experiments.}
\renewcommand{\arraystretch}{1.3}
\setlength{\tabcolsep}{5pt}
\label{tab: dataset}
\begin{tabular}{ccccccccccccc} 
\toprule
Type                & 2014 & 2015 & 2016 & 2017 & 2018 & 2019 & 2020 & 2021 & 2022 & 2023 & 2024 & 2025  \\ 
\midrule
Benign              &   46,536   &   25,279   &   33,752   &   55,951   &  49,651    &  21,360    & 20,650     &   19,586   &   16,484   &    8,804  &   7,182   & 3,402     \\ 
Malware (Seen) &   5,154   &   7,505   & 3,093  &  2,858   &   3,564   &    1,795  &  1,862   &  1,927   &  1,337    &   159   &  336    &   6    \\
Malware (Unknown)      &  263    &   395  &   691   &   4,278   &   4,625   &   44   &   80   &   88   &   81  &   772  &   191   &   58    \\   
Malware (Unseen)    &    /  &     2,838 &   799   &   519   &  422    &   585   &  485    &    473  &   438   &  125   &   302   &  52     \\
                                  
\midrule
Total              &   51,953   &   36,017   &   38,335   &  63,606    &   58,261   &   23,784   &  23,077    &   22,074   &  18,340    &    9,860  &    8,011  &   3,518    \\
\bottomrule
\end{tabular}
\end{table*}

%% file: Table/RQ3.tex
\begin{table}
\centering
\caption{The sum of feature contribution scores (FCS) for the comparison schemes during training and testing in both closed and open-world settings. AG and GR represent the baselines APIGraph~\cite{apigraph} and GuideRetraining~\cite{guide_retraining}.}
\renewcommand{\arraystretch}{1.3}
\setlength{\tabcolsep}{3pt}
\label{tab:rq3}
\begin{tabular}{ccccccccc} 
\toprule
\multirow{2}{*}{\begin{tabular}[c]{@{}c@{}}Normalize \\Dataset No.\end{tabular}} & \multicolumn{4}{c}{Closed world} & \multicolumn{4}{c}{Open world}  \\ 
\cmidrule(lr){2-9}
                  & w/o   & AG    & GR    & Ours      & w/o   & AG    & GR    & Ours    \\ 
\midrule
Train             & 27.22 & 28.33 & 25.15 & 32.15     & 27.22 & 28.33 & 25.15 & 32.15   \\
1                 & 23.98 & 24.49 & 19.04 & 27.98     & 22.91 & 23.62 & 18.58 & 27.33   \\
2                 & 21.70 & 23.53 & 18.44 & 26.14     & 19.46 & 22.57 & 17.04 & 25.74   \\
3                 & 19.78 & 20.27 & 17.12 & 22.92     & 17.56 & 19.29 & 16.10 & 21.97   \\
4                 & 17.94 & 18.04 & 15.11 & 20.16     & 16.68 & 16.74 & 13.72 & 19.26   \\
5                 & 15.73 & 15.82 & 13.52 & 18.72     & 14.69 & 15.09 & 12.06 & 18.57   \\
6                 & 14.26 & 14.23 & 12.46 & 16.63     & 13.56 & 13.75 & 11.10 & 16.59   \\
7                 & 13.14 & 13.21 & 11.67 & 14.89     & 12.12 & 13.19 & 11.63 & 14.75   \\
8                 & 12.58 & 12.47 & 10.88 & 13.83     & 11.75 & 11.83 & 10.05 & 13.66   \\
9                 & 12.26 & 12.31 & 10.52 & 12.91     & 11.24 & 11.12 & 9.30  & 12.38   \\
10                & 11.98 & 12.01 & 6.83  & 12.52     & 10.51 & 10.62 & 6.01  & 11.62   \\
\bottomrule
\end{tabular}
\end{table}

%% file: Table/case.tex
\begin{table*}
\centering
\caption{Top-10 important features for Airpush and Hiddad families from training (Airpush\_tr) and test set (Airpush\_te). Hiddad family (Hiddad\_te) doesn't exist in the training set. The upper part is from \model, and the lower part is from DeepDrebin~\cite{Grossedeepdrebin}. Feature importance rank (1-10) is sorted in descending order of importance score.}
\renewcommand{\arraystretch}{1.3}
\label{tab:case_feature}
\resizebox{\textwidth}{!}{
\begin{tabular}{ccccc} 
\toprule
\textbf{Method}                  & \textbf{Rank} & \textbf{Airpush\_tr}                         & \textbf{Airpush\_te}                                     & \textbf{Hiddad\_te}                                       \\ 
\midrule
\multirow{10}{*}{\model} & 1             & http://schemas\_android\_com/apk/res-auto    & https://play\_google\_com/store/                         & intent\_action\_USER\_PRESENT  \\
& 2             & https://play\_google\_com/store/             & http://schemas\_android\_com/apk/res-auto                & http://schemas\_android\_com/apk/res-auto                 \\
& 3             & http://revmob\_com                           & http://revmob\_com                                       & intent\_action\_PACKAGE\_ADDED                            \\
& 4             & https://android\_revmob\_com                 & https://android\_revmob\_com                             & permission\_ACCESS\_NETWORK\_STATE                        \\
& 5             & permission\_ACCESS\_NETWORK\_STATE           & intent\_action\_PACKAGE\_ADDED                           & http://revmob\_com                                        \\
& 6             & permission\_READ\_PHONE\_STATE               & permission\_ACCESS\_NETWORK\_STATE                       & https://android\_revmob\_com                              \\
& 7             & permission\_ACCESS\_FINE\_LOCATION           & com\_google\_android\_gms\_ads\_AdActivity               & com\_google\_android\_gms\_ads\_AdActivity                \\
& 8             & intent\_action\_USER\_PRESENT                & permission\_ACCESS\_FINE\_LOCATION                       & com\_apperhand\_device\_android\_AndroidSDKProvider       \\
                        & 9             & intent\_action\_PACKAGE\_ADDED               & intent\_action\_USER\_PRESENT                            & permission\_ACCESS\_FINE\_LOCATION                        \\
                        & 10            & permission\_GET\_ACCOUNTS                    & http://xmlpull\_org/v1/doc/features\_html                & getSubscriberId                                           \\ 
\midrule
\multirow{10}{*}{DeepDrebin}    & 1             & http://schemas\_android\_com/apk/res-auto    & https://play\_google\_com/store/                         & http://schemas\_android\_com/apk/res-auto                 \\
                        & 2             & https://play\_google\_com/store/             & http://schemas\_android\_com/apk/res-auto                & intent\_action\_USER\_PRESENT                             \\
                        & 3             & permission\_ACCESS\_FINE\_LOCATION           & android/location/LocationManager;-requestLocationUpdates & intent\_action\_PACKAGE\_ADDED                            \\
                        & 4             & android/accounts/AccountManager;-getAccounts & http://xmlpull\_org/v1/doc/features\_html                & android/location/LocationManager;-requestLocationUpdates  \\
                        & 5             & http://revmob\_com\_                         & permission\_ACCESS\_FINE\_LOCATION                       & permission\_ACCESS\_WIFI\_STATE                           \\
                        & 6             & permission\_GET\_ACCOUNTS                    & https://market\_android\_com                             & android/app/ActivityManager;-getRunningTasks              \\
                        & 7             & permission\_READ\_HISTORY\_BOOKMARKS         & intent\_action\_PACKAGE\_ADDED                           & permission\_ACCESS\_FINE\_LOCATION                        \\
                        & 8             & permission\_ACCESS\_WIFI\_STATE              & android/app/ActivityManager;-getRunningTasks             & permission\_SYSTEM\_ALERT\_WINDOW                         \\
                        & 9             & https://android\_revmob\_com                 & permission\_ACCESS\_WIFI\_STATE                          & http://xmlpull\_org/v1/doc/features\_html                 \\
                        & 10            & permission\_READ\_PHONE\_STATE               & http://revmob\_com                                       & printStackTrace                                           \\
\bottomrule
\end{tabular}}
\end{table*}

%% file: Table/overhead.tex
\begin{table}
\setlength{\tabcolsep}{3pt}
\centering
\caption{Average per-epoch training time (seconds) of robustness-enhancement methods under a fixed batch size for each feature space. w/o indicates the baseline without robustness enhancement; for Drebin, this corresponds to DeepDrebin. \model's training time combines two stages.}
\renewcommand{\arraystretch}{1.3}
\label{tab:overhead}
\begin{tabular}{ccccc} 
\toprule
Feature Space & w/o & APIGraph~\cite{apigraph} & Guide Retraining~\cite{guide_retraining} & Ours  \\ 
\midrule
Drebin~\cite{Grossedeepdrebin}        &   1.211  &   1.395      &      5.734            &    3.764   \\
Malscan~\cite{malscan}       &    1.582  &     /     &          7.750    &    5.280   \\
BERTroid~\cite{bertroid}     &   655.532  &     /     &         1408.017         &   1160.588    \\
\bottomrule
\end{tabular}
\end{table}

%% file: Table/Env.tex
\begin{table}
\centering
\caption{AUT (F1,12m) of the DeepDrebin~\cite{Grossedeepdrebin} under different temporal environment segmentation methods.}
\renewcommand{\arraystretch}{1.3}
\setlength{\tabcolsep}{5pt}
\label{tab: env}
\begin{tabular}{ccccccc} 
\toprule
\multirow{2}{*}{\begin{tabular}[c]{@{}c@{}}Test\\Years\end{tabular}} & \multicolumn{5}{c}{DeepDrebin~\cite{Grossedeepdrebin}}                        & \multicolumn{1}{l}{}  \\ 
\cmidrule(lr){2-7}
                  & w/o   & monthly       & quarterly     & n=4   & n=8   & n=12                  \\ 
\midrule
2015              & 0.859 & \uline{0.928} & 0.915         & 0.914 & 0.902 & 0.905                 \\
2016              & 0.813 & \uline{0.879} & 0.866         & 0.864 & 0.854 & 0.853                 \\
2017              & 0.752 & \uline{0.816} & 0.801         & 0.800 & 0.783 & 0.799                 \\
2018              & 0.726 & \uline{0.794} & 0.783         & 0.780 & 0.758 & 0.754                 \\
2019              & 0.713 & 0.769         & \uline{0.775} & 0.762 & 0.745 & 0.741                 \\
2020              & 0.700 & 0.732         & \uline{0.764} & 0.730 & 0.718 & 0.719                 \\
2021              & 0.689 & 0.717         & \uline{0.750} & 0.709 & 0.701 & 0.705                 \\
2022              & 0.661 & 0.688         & \uline{0.717} & 0.667 & 0.669 & 0.671                 \\
2023              & 0.638 & 0.665         & \uline{0.690} & 0.634 & 0.643 & 0.645                 \\
\bottomrule
\end{tabular}
\end{table}

%% file: Table/Regularization.tex

\begin{table}
\centering
\caption{AUT (F1,12m) of the Deepdrebin~\cite{Grossedeepdrebin} under different regularization methods.}
\renewcommand{\arraystretch}{1.3}
\label{tab: regularization}
\begin{tabular}{cccccc} 
\toprule
\multirow{2}{*}{} & \multicolumn{5}{c}{DeepDrebin}                                                               \\ 
\cmidrule(lr){2-6}
                  & ERM   & \begin{tabular}[c]{@{}c@{}}Early\\Stopping\end{tabular} & Dropout & $\ell$2 & Ours   \\ 
\midrule
2015              & 0.856 & 0.833                                                   & 0.859   & 0.869   & 0.928  \\
2016              & 0.808 & 0.795                                                   & 0.818   & 0.822   & 0.877  \\
2017              & 0.742 & 0.736                                                   & 0.754   & 0.759   & 0.815  \\
2018              & 0.703 & 0.701                                                   & 0.726   & 0.727   & 0.794  \\
2019              & 0.682 & 0.682                                                   & 0.713   & 0.704   & 0.769  \\
2020              & 0.664 & 0.668                                                   & 0.696   & 0.681   & 0.732  \\
2021              & 0.651 & 0.657                                                   & 0.686   & 0.669   & 0.717  \\
2022              & 0.628 & 0.635                                                   & 0.658   & 0.641   & 0.688  \\
2023              & 0.611 & 0.617                                                   & 0.631   & 0.619   & 0.665  \\
\bottomrule
\end{tabular}
\end{table}